\documentclass[twocolumn,amsmath,amssymb,floatfix, superscriptaddress,prb]{revtex4-2}
\usepackage{lineno,hyperref}
\usepackage{bm}
\usepackage{amsmath}
\usepackage{gensymb}
\usepackage{epstopdf}
\usepackage{booktabs}
\usepackage{longtable}
\usepackage{tabularx}
\usepackage{multirow}
\usepackage{setspace}
\usepackage{mathrsfs}

\usepackage{graphicx}
\usepackage{ulem}
\usepackage{xcolor}
\usepackage{hyperref}

\begin{document}

\title{Origin of magnetic ordering in half-Heusler RuMnGa}

\author{Sudip Chakraborty}
\email{sudip698@gmail.com}
\affiliation{Condensed Matter Physics Division, Saha Institute of Nuclear Physics,
              1/AF Bidhannagar, Kolkata 700064, India}
\affiliation{Homi Bhabha National Institute, Anushaktinagar, Mumbai 400094, India}
\author{Shuvankar Gupta}
\affiliation{Condensed Matter Physics Division, Saha Institute of Nuclear Physics,
               1/AF Bidhannagar, Kolkata 700064, India}
\affiliation{Homi Bhabha National Institute, Anushaktinagar, Mumbai 400094, India}
\author{Santanu Pakhira}
\affiliation{Ames National Laboratory, Iowa State University, Ames, Iowa 50011, USA}
\author{Anis Biswas}
\affiliation{Ames National Laboratory, Iowa State University, Ames, Iowa 50011, USA}
\author{Yaroslav Mudryk}
\affiliation{Ames National Laboratory, Iowa State University, Ames, Iowa 50011, USA}
\author{Renu Choudhary}
\affiliation{Ames National Laboratory, Iowa State University, Ames, Iowa 50011, USA}
\author{Amit Kumar}
\affiliation{Homi Bhabha National Institute, Anushaktinagar, Mumbai 400094, India}
\affiliation{Solid State Physics Division, Bhabha Atomic Research Centre, Mumbai 400094, Maharashtra,  India}
\author{Amitabh Das}
\affiliation{Homi Bhabha National Institute, Anushaktinagar, Mumbai 400094, India}
\affiliation{Solid State Physics Division, Bhabha Atomic Research Centre, Mumbai 400094, Maharashtra,  India}
\author{Chandan Mazumdar}

\affiliation{Condensed Matter Physics Division, Saha Institute of Nuclear Physics,
               1/AF Bidhannagar, Kolkata 700064, India}
\affiliation{Homi Bhabha National Institute, Anushaktinagar, Mumbai 400094, India}

\date{\today}

\begin{abstract}
The half-Heusler alloy RuMnGa having valence electron count (VEC) 18 has recently been theoretically proposed to exhibit compensated ferrimagnetic (CFiM) character instead of the expected nonmagnetic ground state. On the other hand, a preliminary experimental study proposed  ferromagnetic (FM) ordering. As no half-Heusler system with VEC 18 is known to exhibit magnetic ordering, we have investigated the details of crystal structure and magnetic properties of RuMnGa using a combination of experimental tools, viz., x-ray and neutron diffraction techniques, dc and ac susceptibility, isothermal magnetisation, heat capacity, resistivity and neutron depolarisation measurements. Rietveld refinements of x-ray and neutron diffraction data suggest single phase nature of the compound with elemental composition RuMn$_{0.86}$Ga$_{1.14}$.  We have shown that the system exhibits FM-type ordering owing to the inherent presence of this minor off-stoichiometry, showing very low magnetic moment. The system also exhibits reentrant canonical spin-glass behaviour, which is rarely observed in half-Heusler alloys. The temperature coefficient of resistivity changes its sign from negative to positive and further to negative as the temperature decreases.
\end{abstract}

\maketitle

\section{Introduction}

The Heusler alloys and their derivative compounds continue to attract considerable attention of the condensed matter physics and materials science communities due to a plethora of tailor-made properties that are both fundamentally interesting and potentially functional. Well known examples include half-metallic ferromagnetism (HMF), ferromagnetic shape memory effects, unusual thermoelectricity, giant magnetocaloric effect, and formation of many topological states, viz., magnetic skyrmions, Weyl semimetals, and others~\cite{katsnelson2008half,de1983new,mondal2018ferromagnetically,devi2019improved,aksoy2009magnetic,planes2009magnetocaloric,wang2016time,venkateswara2019coexistence,meshcheriakova2014large,zuo2018zero,gupta2022coexisting}. Generally, Heusler phases are classified either as full-Heusler, commonly represented by the idealized X$_2$YZ stoichiometries, where X and Y are transition elements and Z is main-group element, or half-Heulser, often quoted as XYZ compounds. Structurally ordered full-Heusler alloys crystallize with the L2$_1$-type structure (space group: \textit{Fm$\bar{3}$m}, No. 225) that consists of four interpenetrating face-centered cubic (fcc) sub-lattices. For the case of half-Heusler alloy, one of these fcc sub-lattice remains vacant and it crystalizes in Y-type of structure (space group: \textit{F$\bar{4}$3m}, No. 216). A rather simple cubic crystal structure makes them an ideal model system for fundamental understanding of \textit{d}-band magnetism and magneto-transport phenomena~\cite{graf2011simple}. In the field of spintronics, the high spin-polarization property of HFM Heusler alloys due to unique band structures, in which one spin channel is metallic, whereas the other spin channel is semi-conducting in nature, can bring future technological advancements in high density data storage.

The ferromagnetic spin structure of the HMF, however, poses a flip side on the device performance due to the inherent presence of large dipole field. To get rid of these stray fields, a significant attention is currently invested in zero magnetic moment spintronics, as these systems are devoid of any intrinsic dipole fields and are exceptionally stable against externally applied magnetic fields~\cite{park2011spin,baltz2018antiferromagnetic,wang2012room}. Compensated ferrimagnets (CFiM) thus turn out to be an ideal solution. They have a net zero magnetic moment and unique band structure conducive to spin polarized currents. An important feature of CFiMs is that they do not create any external magnetic field and hence are devoid of the magnetostatic energy. Thus, CFiMs are considered as ideal candidates for achieving 100\% spin-polarised current without any net magnetic moment, facilitating novel spintronic devices~\cite{cai2020ultrafast,nayak2015design,geprags2016origin}.

In cases of  HAs, the net magnetic moment present in the samples generally follows the Slater-Pauling (SP) rule, according to which, the  expected saturation magnetization ($M_{sat}$) of a full-HA is $M_{sat}$ = ($N_V$-24)$\mu_{\rm B}$/f.u., whereas for a half-HA it is $M_{sat}$ = ($N_V$-18)$\mu_{\rm B}$/f.u., with $N_V$ being the number of valence electrons~\cite{graf2011simple}.

Thus, according to SP rules, the half-HA with valence electron count (VEC) of 18  or full-HA with VEC 24 would have equal contribution in the spin-up and spin-down bands making those  systems paramagnetic. On the other hands, HAs with  VEC other than those particular numbers (18 for half and 24 for full HA) must have asymmetrically populated spin-up and spin-down bands  generally giving rise to ferromagnetic (FM) ordering although a very few HAs also show antiferromagnetic (AFM) ordering. Interestingly,  a number of recent theoretical computational works predict that the CFiM structure can also be stabilized in cases of full-HA with VEC equals to 24 and half-HA with VEC equals to 18, which leads to a new possibility of discovering more efficient HA for spintronics application~\cite{galanakis2007ab,shi2018prediction,patel2018first,ma2017computational}. However, that theoretical prediction is yet to be conclusively verified with experiments and an open question still remains as whether CFiM half metallic HA is practically realized or not~\cite{ma2017computational}.

RuMnGa is one such promising HA compound forming in C1$_b$ crystal structure and theoretically proposed to exhibit a very low moment FiM characteristic, where all the constituent elements have nearly equal magnetic contribution. Interestingly, a few decades old work instead reported the compound to exhibit low-moment ferromagnetism (0.30$\mu_{\rm B}$/mole) with rather high paramagnetic Curie-Weiss temperature ($\theta_p$ = 220 K)~\cite{hames1971ferromagnetism}. No further details of the experimental result were provided. However, the experimentally observed lattice parameter, also considering the C1$_b$ crystal structure, was found to be larger than the theoretically predicted one for which FiM ground state was proposed.  Interestingly, the formation energy estimated using this theoretically derived lattice parameter turns out to be positive for RuMnGa~\cite{ma2017computational}, raising suspicion on the formation of the material in this structure. The contradictory nature of theoretical studies and experimental results on RuMnGa is thus quite intriguing.

In this work, we have investigated the structural properties of RuMnGa compound using x-ray and neutron diffraction techniques, whereas the magnetic ground state and physical properties of the system are studied through dc and ac magnetization, neutron diffraction, neutron depolarization as well as thermal- and electrical-transport measurements. Attempts have been made to explain these features.

\section{Experimental details}

Polycrystalline samples of RuMnGa ($\sim$ 8 g) were synthesized by standard arc melting technique. Highly pure ($>$99.9 wt.\%) raw elements were melted in argon atmosphere on a water cooled Cu hearth. To compensate the weight loss due to evaporation, Mn and Ga were taken with 2\% in excess weight. Better homogeneity of the sample was ensured by flipping over and re-melting the ingot 5 times. Crystallographic structure and phase purity were determined from the x-ray diffraction (XRD) spectrum measured on a powdered sample at room temperature, using a commercial diffractometer (rotating anode, 9 kW, Model: TTRAX-III, Rigaku Corp., Japan) using Cu-K$\alpha$ radiation ($\lambda$ = 1.54056 \AA). Full Rietveld refinement of the XRD data was performed with the aid of the Fullprof software package~\cite{rodriguez1993recent}.

Magnetic properties were studied in a commercial SQUID magnetometer (Quantum Design Inc., USA) in the temperature region 2--380 K and in the magnetic field region 0--70 kOe. Magnetic susceptibility measurements were performed in standard zero-field-cooled (ZFC) and field-cooled (FC) methods as described in~\cite{chakraborty2022ground}. The ac susceptibility measurements were performed in the temperature region 5--300 K at different frequencies with an excitation ac magnetic field of 6 Oe during warming cycle after cooling down the sample to 5 K in the absence of field. Powder Neutron Diffraction (PND) was performed in PD2 powder neutron diffractometer ($\lambda$ = 1.24395  \AA) in Dhruva reactor, Bhaba Atomic Research Centre, Mumbai, India in aluminum sample holder. Neutron depolarization measurements were performed on the Polarized Neutron Spectrometer at the Dhruva reactor. The incident neutron beam was polarized using a monochromator-cum-polarizer single crystal of Cu$_2$MnAl (111) with polarization efficiency $\sim$ 99\% as measured by the flipping ratio method. The polarization of the neutron beam along its path was preserved by a vertical guide field of 110 Oe. A dc flipper before sample was used to invert the polarization of the neutron beam. The polarization of the transmitted neutron beam was analyzed by Co$_{0.92}$Fe$_{0.08}$ (200) single crystal.

Heat capacity of the sample was measured using relaxation technique in Physical Property Measurement System (PPMS, Quantum Design Inc., USA), in the temperature range 2--300 K. Electrical resistivity measurement was performed in standard four probe method in a PPMS set-up. Silver epoxy was used to make contact on the rectangular shaped polished sample of dimension (2.3 mm $\times$ 1.1 mm $\times$ 0.49 mm).

\section{Results and Discussions}
\subsection {Structural Characterization}
\label{sec:structure}

\begin{figure}[h]
\centering
\includegraphics [width=0.5\textwidth]{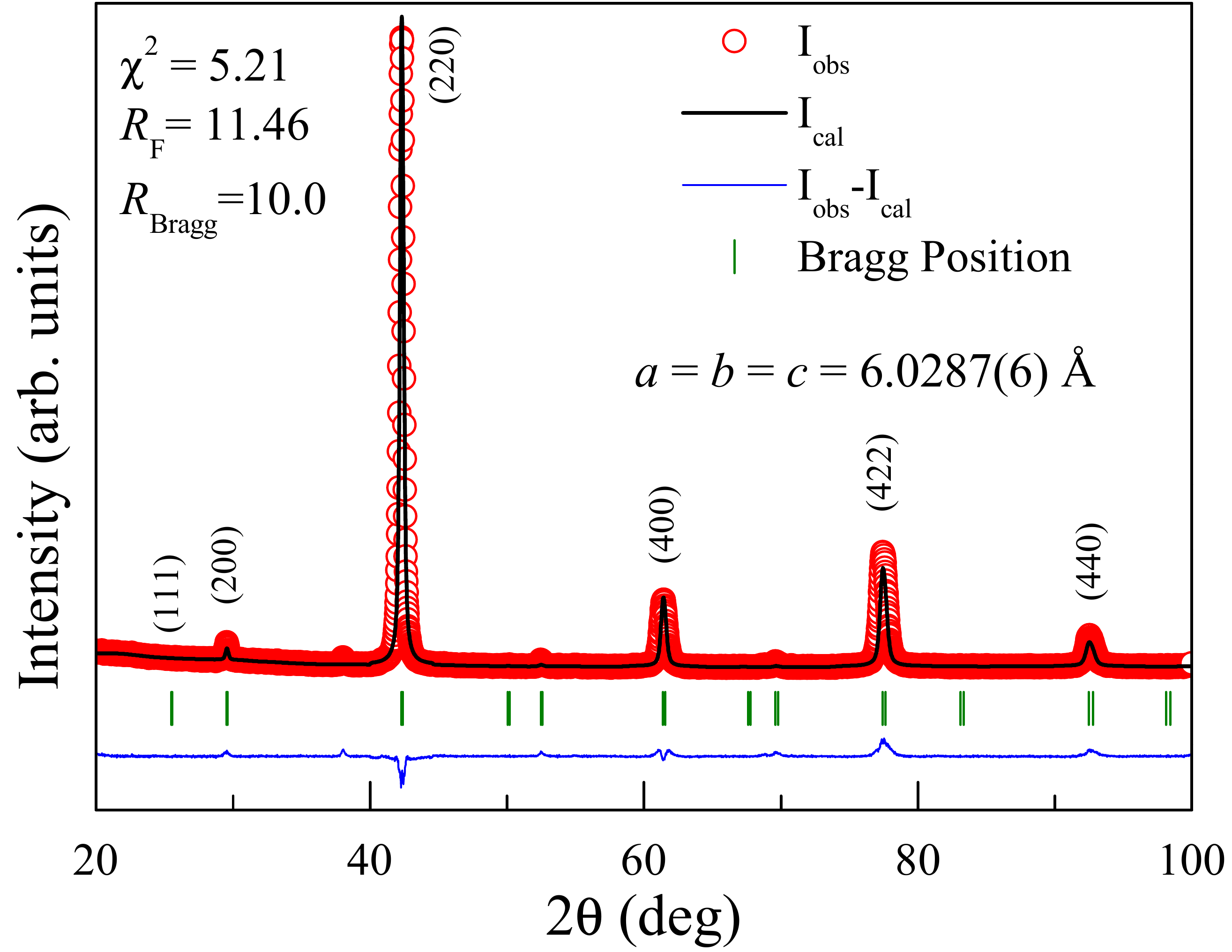}
\caption{(a) Room-temperature XRD data along with full profile Rietveld refinement. The red open circle denotes experimental data, black line is the calculated data and blue line represents difference between experimental and calculated data, whereas, the olive vertical lines indicate the positions of the Bragg peaks.}
\label{fig:XRD}
\end{figure}

Full Rietveld analysis of the room-temperature powder XRD spectrum shown in Fig.~\ref{fig:XRD} reveals that \mbox{RuMnGa} forms essentially in single phase in the space group $F\bar{4}3m$ (No. 216).
The lattice parameter obtained from this analysis is 6.0287(6) \AA, matching quite well with the earlier experimental report~\cite{hames1971ferromagnetism}, but differs greatly with the value proposed through the theoretical calculations~\cite{ma2017computational}. Crystallographically well-ordered half HAs belonging to the MgAgAs-type (commonly called C1$_b$-type) are represented stoichiometrically by XYZ, where X, Y are the transition elements and Z is the $p$ block elements. In this crystal structure, X, Y and Z atoms occupy 4\textit{a}, 4\textit{b} and 4\textit{c} positions respectively, in space group \textit{F$\bar{4}$3m} (No. 216). When an additional atom is introduced, as in quaternary HA, that atom occupies the 4$d$ position. However, in reality, most of the HAs tend to form with structural disorders, mostly in A2 or B2 type~\cite{graf2011simple}. In B2 type disorder, atoms in 4$a$/4$b$ sites and 4$c$/4$d$ sites got intermixed, whereas in A2 type disorder, atoms in all the 4 sites are randomly distributed. Such structural disorder in HAs can be identified and distinguished by examining the (111) and (200) superlattice reflection peaks in the diffraction pattern. In B2 type disorder, (111) Bragg peak is always absent whereas in A2 type disorder, both (111) and (200) peaks are absent. In the earlier report the crystal structure of RuMnGa was claimed to be the conventional C1$_b$ structure, without providing any details of structural analysis~\cite{hames1971ferromagnetism}. The absence of (111) Bragg peak in the presented XRD pattern (Fig.~\ref{fig:XRD}) clearly indicates the departure from the perfectly ordered C1$_b$-type structure, suggesting instead the B2 type disordered structure. The B2-type disorder is also confirmed by the neutron diffraction (ND) pattern (discussed later in Sec.~\ref{sec:ND}) where  the (111) peak also remains absent. The compositional analysis carried out through combined application of full Rietveld refinement of the ND (discussed later in Sec.~\ref{sec:ND}) and XRD data suggest that the Mn and Ga atoms are almost equally intermixed between 4$a$ and 4$b$ sites, while Ru atoms are placed in 4$c$ and 4$d$ sites with nearly 50\% occupancy at both the sites. During the fitting process, Ru occupancies are kept fixed, and occupancies of all other elements are allowed to vary. We found that both the XRD and ND spectra (Sec.~\ref{sec:ND}) can be described with a single structural model (Table~\ref{tab:my-table}) having the sample composition of RuMn$_{0.86}$Ga$_{1.14}$.

\begin{table}[]
\caption{Crystallographic parameters of RuMnGa obtained from combined full Rietveld refinement of room temperature powder XRD and ND data.}

\begin{tabular}{ccccccccc}
\toprule
\hline
\multicolumn{2}{c}{Compound}                                      & \multicolumn{6}{c}{RuMnGa}                                            \\ \midrule
\multicolumn{2}{c}{Crystal structure}                             & \multicolumn{6}{c}{MgAgAs type}                                       \\
\multicolumn{2}{c}{Space group}                                   & \multicolumn{6}{c}{{\textit{F$\bar{4}$3m}}  (No. 216)} \\
\multicolumn{2}{c}{$a$ = $b$ = $c$ =}                                   & \multicolumn{6}{c}{6.0287(6) \AA }                     \\  \midrule
                                                                                                   \\
\hline
Atom & \begin{tabular}[c]{@{}c@{}}Wyckoff\\ position\end{tabular} & x          &\hspace{0.5 cm}     & y         &\hspace{0.5 cm}     & z     &\hspace{0.5 cm}        & occupancy          \\
\hline
\hline
Ru1  & 4$c$                                                         & 0.25     &       & 0.25    &       & 0.25     &     & 0.50             \\
Ru2  & 4$d$                                                         & 0.75     &       & 0.75    &       & 0.75     &     & 0.50             \\
Mn1  & 4$a$                                                         & 0        &       & 0       &       & 0        &     & 0.41(1)          \\
Mn2  & 4$b$                                                         & 0.50     &       & 0.50    &       & 0.50     &     & 0.45(1)          \\
Ga1  & 4$a$                                                         & 0        &       & 0       &       & 0        &     & 0.59(1)          \\
Ga2  & 4$b$                                                         & 0.50     &       & 0.50    &       & 0.50     &     & 0.54(1)          \\ \bottomrule
\end{tabular}

\label{tab:my-table}
\end{table}

\subsection {dc Magnetization Study}
\label{sec:DC magnetization}
\begin{figure}[ht]
\centering
\includegraphics [width=0.5\textwidth]{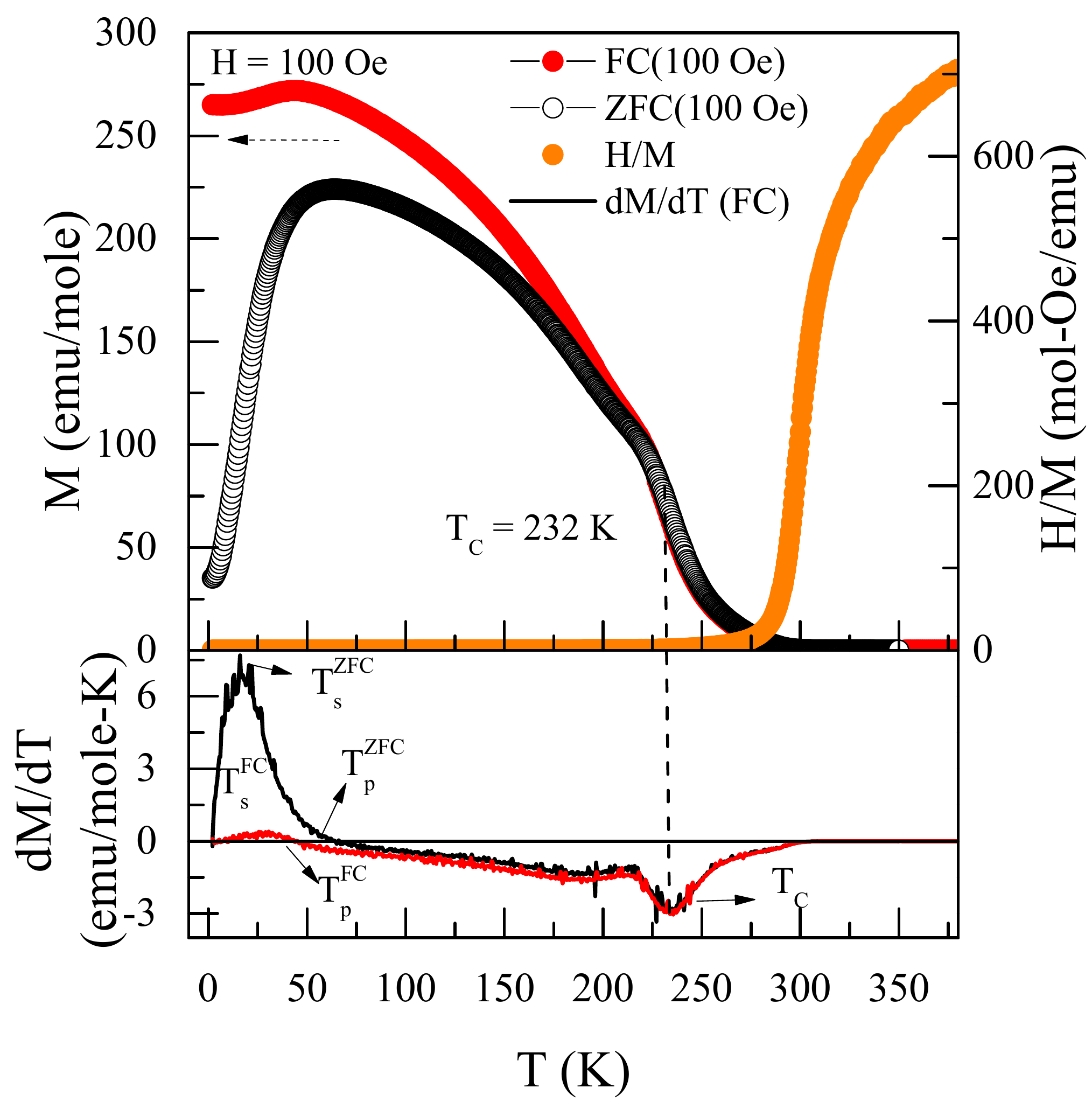}
\caption{(Upper panel) Magnetization measured in ZFC and FC protocol  vs. Temperature plot at 100 Oe applied
magnetic field (Left scale) and inverse susceptibility of the data measured using the FC protocol (Right scale). In the lower panel the temperature dependence of $dM/dT$ is plotted, the different transitions are marked accordingly.}
\label{fig:MT}
\end{figure}

The temperature variation of magnetization for $H = 100$~Oe under both ZFC and FC conditions is presented in Fig.~\ref{fig:MT} in the temperature range 2--380~K\@. As observed in the $M(T)$ and $dM(T)/dT$ behavior, the system undergoes a magnetic transition below $T_{\rm C} = 232$~K, which was earlier described as a ferromagnetic transition according to a previous study~\cite{hames1971ferromagnetism}.  Additionally, two more distinct low-temperature anomalies ($T_p$ and $T_s$) are observed in the magnetization behavior, as shown in Fig.~\ref{fig:MT}. The temperatures corresponding to those anomalies slightly differ in the ZFC and FC measurements. The ZFC magnetization exhibits a broad peak around $T_p^{\rm ZFC}$ $\sim$ 65 K, whereas the corresponding FC anomaly is observed at $T_p^{\rm FC}$ $\sim$ 45 K. A thermal hysteresis between the ZFC and FC magnetization curves is present below $T_{\rm C}$, as found in many FM/FiM systems for low field susceptibility measurement. As these transitions are rather broad in nature, the peak temperature is determined as the temperature where $M(T)$ behavior changes the slope yielding a negative to positive crossover in the corresponding $dM(T)/dT$ curve (lower panel of Fig.~\ref{fig:MT}). The decrease in $M(T)$ below $T_p$ in both ZFC and FC modes suggest the development of anti-parallel spin arrangement or a glassy magnetic phase below that temperature. At further low temperature, an additional weak anomaly is observed in the $M(T)$ behavior, where the ZFC and FC magnetization exhibits a weak inflection with decreasing temperature below $T_{s}^{\rm ZFC}$ $\sim $ 20 K and $T_{s}^{\rm FC}$ $\sim$ 30 K, respectively, as determined from the corresponding $dM(T)/dT$ behavior. However, as the change in magnetisation is rather hard to discern, no conclusion can be drawn on its origin from the dc susceptibility. The feature, on the other hand, has also cast its signature in the isothermal magnetisation results (see below) and found prominence in the ac susceptibility measurements (Sec.~\ref{sec:AC susceptibility}). However, multiple explanations could be provided for the reduction in dc magnetic susceptibility below $T_p$, both in the ZFC as well as FC measurements : (i) one may consider the system to be ferrimagnetic, as proposed from the theoretical study~\cite{ma2017computational}, where the relative competition of multiple magnetic sublattices could result in such reduction in magnetic moment at low temperature; (ii) The system can also order ferromagnetically, where the transition at $T_p$ could be explained as due to spin reorientation; (iii) One may also have the development of reentrant spin-/cluster-glass feature below $T_p$, where the magnetically ordered phase may completely or even partially transformed into the magnetically glassy phase. The inverse susceptibility curve (shown in right panel of Fig.~\ref{fig:MT}), deviates from linearity below $T \sim 323$~K which is quite higher than the ordering temperature $T_{\rm C}$, and is typically observed in different ferrimagnetic systems (Fig.~\ref{fig:MT}). However, such feature can also be explained using a ferromagnetic model, where the presence of magnetic precursor effect, i.e., development of short range magnetic clusters above $T_{\rm C}$ would also result in similar characteristics.

\begin{figure}[ht]
\centering
\includegraphics [width=0.5\textwidth]{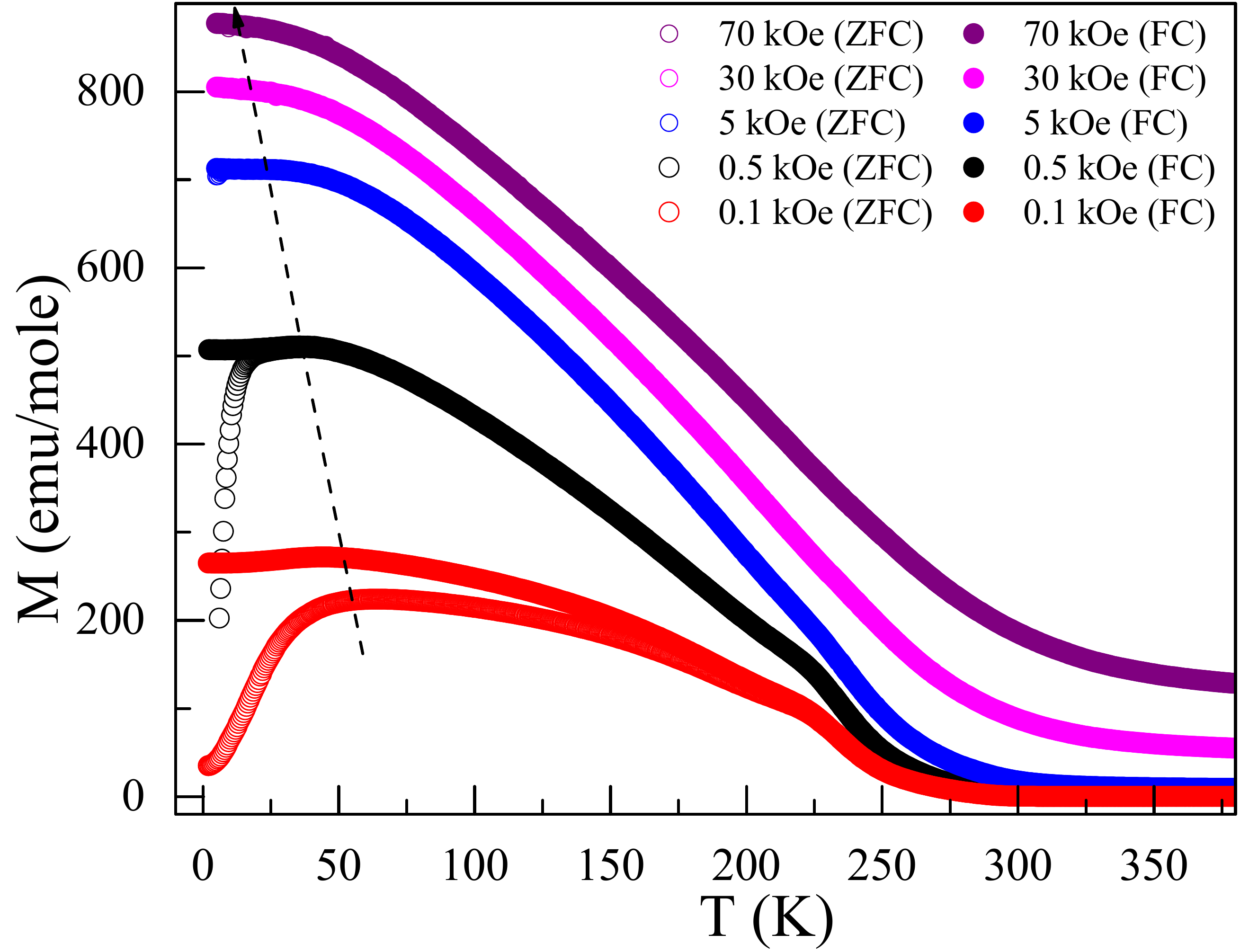}
\caption{Magnetization vs Temperature plot at various applied magnetic field. The arrow shows the gradual shift of the $T_p$ with the applied external field.}
\label{fig:FullMT}
\end{figure}

To have a better understanding of the magnetic transitions in this compound, magnetization measurements were carried out at different magnetic fields, as shown in Fig.~\ref{fig:FullMT}. It can be seen that the magnetic irreversibility above $T_p$ disappears as the magnetic field is enhanced to 500 Oe, whereas the irreversibility below $T_p$ vanishes at a moderate field of 5 kOe. Thus, the weak thermal hysteresis above $T_p$ could be due to the weak pinning of domain walls~\cite{mazumdar2000smni}, whereas the relatively stronger thermal hysteresis below $T_p$ could be due to additional glassy-like state formation in the system.

\begin{figure}[ht]
\centering
\includegraphics [width=0.5\textwidth]{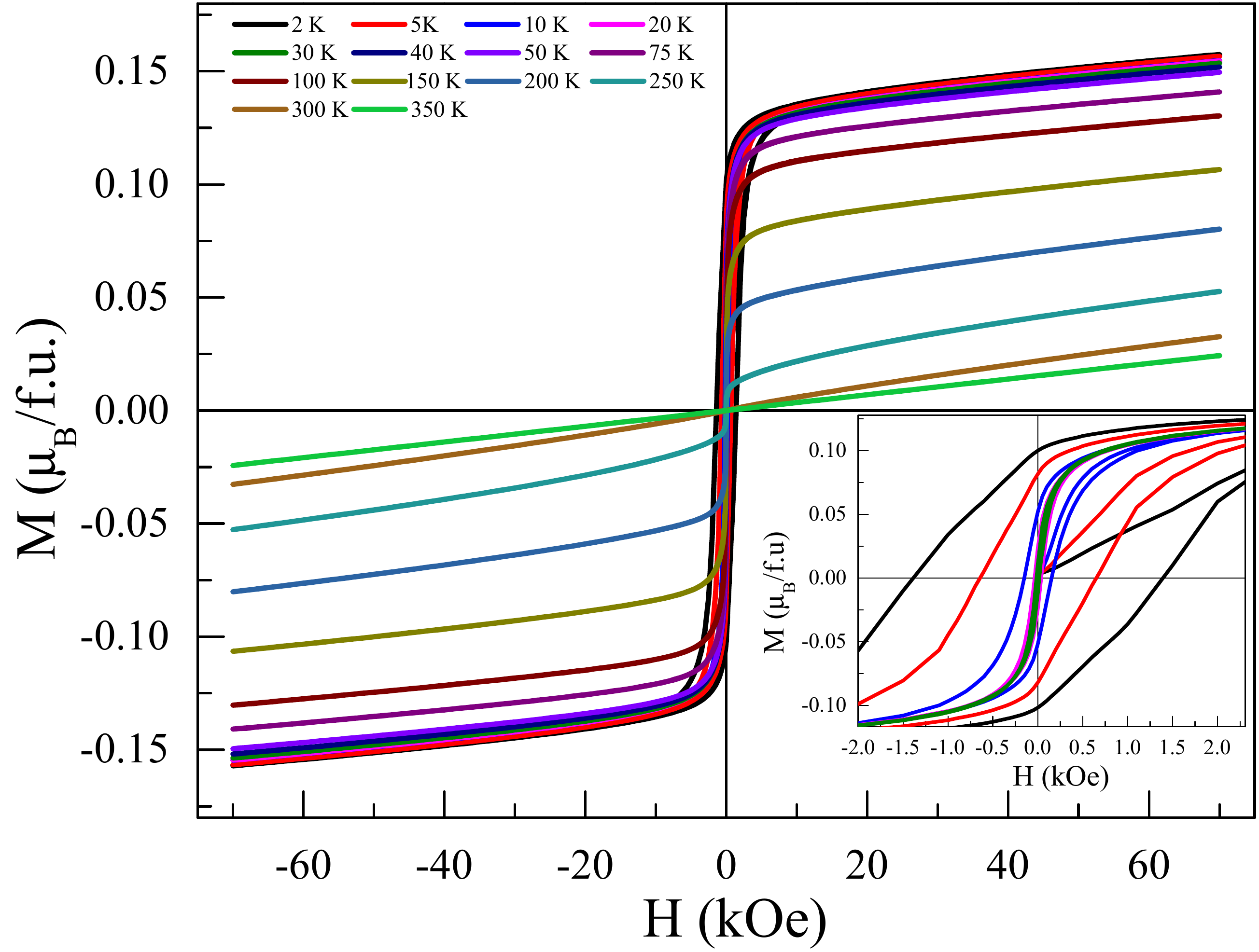}
\caption{Magnetization vs. applied magnetic field plot at various temperature.}
\label{fig:MHFull}
\end{figure}

In order to understand the magnetic field dependence of the system, isothermal magnetization $M(H)$ measurements have been carried out at different temperatures near different phase transitions (Fig.~\ref{fig:MHFull}). The $M(H)$ behavior exhibits a soft FM/FiM-like pattern for $T < 250$~K. As the temperature decreases, the overall $M(H)$ response increases with decreasing temperature down to 75 K, but remain apparently unaltered for $T < T_p$. Even at the lowest temperature (2 K) and highest applied field (70 kOe), the magnetization does not saturate, exhibiting a weak field dependence. Such type of behavior is generally observed in glassy systems below the spin-freezing temperature~\cite{kroder2019spin}. The spontaneous value of magnetic moment extrapolated to $H$ = 0 at 2 K has been estimated to be 0.12 $\mu_{\rm B}$/f.u., closely matching both the theoretical prediction for FiM configuration and previous experimental results revealing the FM type spin ordering.  Incidentally, if we consider the experimentally determined elemental composition from the Rietveld refinement of XRD and ND data for the material to be RuMn$_{0.86}$Ga$_{1.14}$, the VEC turn out to be 17.64, instead of 18 expected for the full stoichiometric composition. Considering the actual composition of the sample the SP rule applicable for a ferromagnetic system suggests a magnetic moment of about 0.36 $\mu_{\rm B}$/f.u., which is also close to the experimentally observed saturation moment. Thus, if one assigns the small value of magnetic moment in this system to the compositional off-stoichiometry, the magnetic ordering at $T_{\rm C}$ could be of ferromagnetic type. The above mentioned magnetisation result, however, could not distinguish between the two different possible magnetic ground states in ideal VEC 18 compound, viz., PM or CFiM. Although the structural vacancies could result in FM spin arrangement, as described above, the probability of the development of small moment FiM spin arrangement can not be completely ruled out either without the neutron diffraction studies. Additionally, weak magnetic hysteresis is also observed at lower temperatures for $T < T_{s}$, below which magnetic coercivity increases with decreasing temperature. This is in accordance with the field-dependent magnetic susceptibility behavior discussed above and also with the ac-susceptibility results, discussed in Sec.~\ref{sec:ND}. Despite the FM/FiM ordering below $T_{\rm C}$, a change in the nature of magnetic spin arrangement is evident below $T_s$, suggesting the development of another magnetic phase, replacing or coexisting with the FM/FiM ordering.

\subsection {Neutron Diffraction}
\label{sec:ND}
\begin{figure}[h]
\centering
\includegraphics [width=0.5\textwidth]{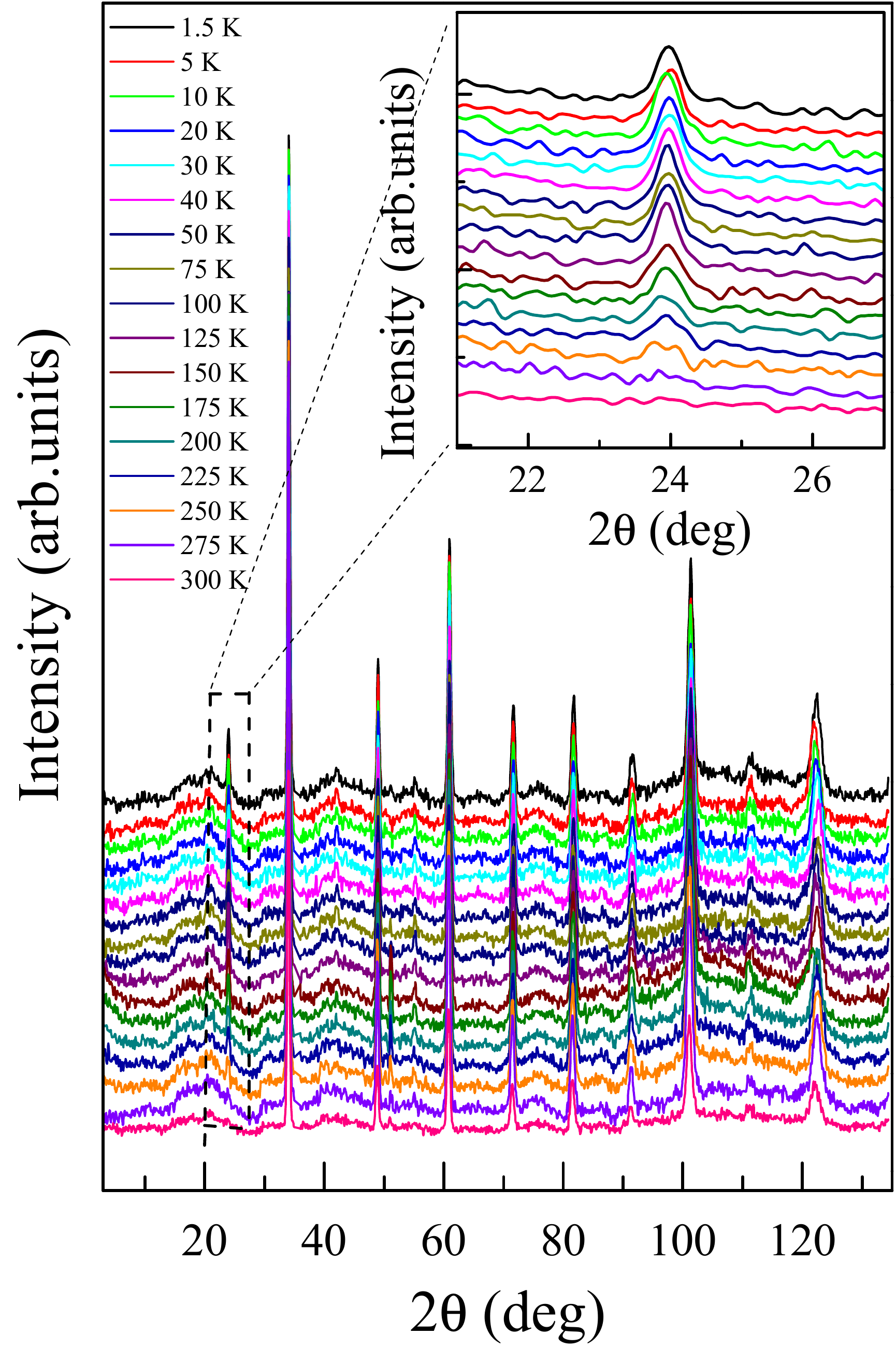}
\caption{Neutron Diffractions pattern of RuMnGa taken at various temperatures ranging from 1.5--300~K. The gradual evolution of (200) peak is shown in the inset.}
\label{fig:Neutron Full}
\end{figure}

\begin{figure}[t]
\centering
\includegraphics [width=0.5\textwidth]{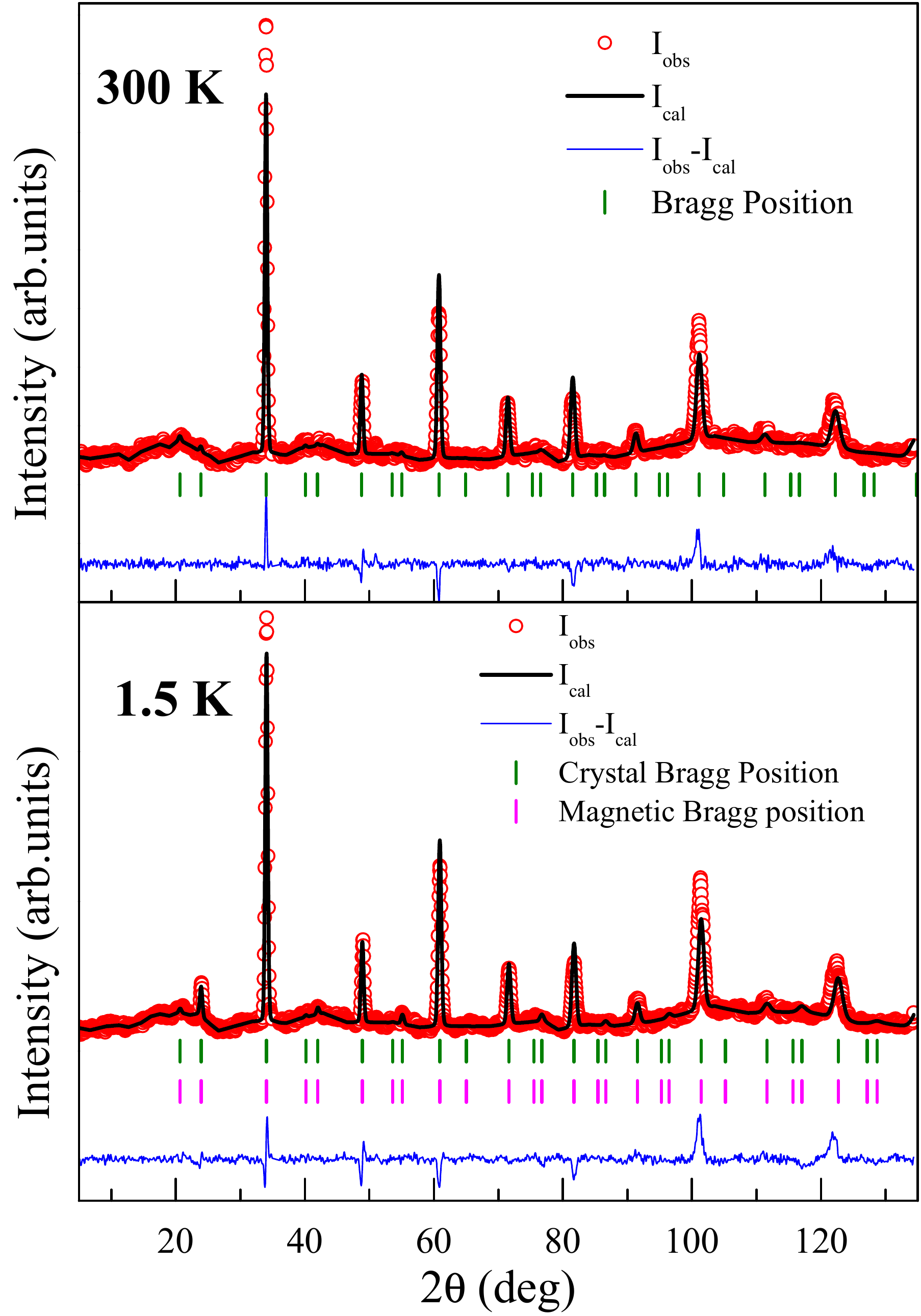}
\caption{Neutron Diffraction pattern along with full Rietveld refinement at temperatures 300 K (top) and 1.5 K (bottom). Recorded data are represented by red open circles, the calculated are shown as a black line, the difference between theoretical and experimental results by blue line and the crystallographic Bragg peak positions are represented by olive lines. Magenta lines represent magnetic Bragg peaks.}
\label{fig:Neutron Fit}
\end{figure}

Zero-field neutron diffraction (ND) measurements of RuMnGa have been carried out due to this method's capability to distinguish between the theoretically predicted FiM structure and experimentally observed FM one. The ND measurements were performed on powdered sample at various temperatures well above and below the magnetic transitions. As mentioned earlier (Sec.~\ref{sec:structure}), Rietveld refinement of the ND pattern at 300 K helped us to understand the atomic distribution between the sites more precisely. Rietveld refinements of all the data taken above $T_{\rm C}$ were performed considering {\textit{F$\bar{4}$3m}} space group by varying the fractional occupancies while restricting the total occupancies of each sites to maximum value of 1.  The ND data was also collected at several temperatures below $T_{\rm C}$, in order to understand the magnetic spin arrangement. None of these patterns show any additional peaks other than those corresponds to the \textit{F$\bar{4}$3m} space group and thus rules out the AFM nature of the magnetic ordering. The long range nature of magnetic ordering however is reflected in the ND pattern of 1.5 K where the (200) peak exhibits clear enhancement in intensity. A close investigation reveals that the intensity of (200) peak starts to increase below $T_{\rm C}$, whereas no change in pattern could be detected within the resolution limit of the instrument across $T_s$ and $T_p$ (Inset: Fig.~\ref{fig:Neutron Full}). This  feature of ND data suggests that the nature of magnetic order essentially remains unaltered in the whole temperature range from 1.5 to $T_{\rm C}$. Although both the FM as well as FiM spin arrangement is in consonance with the enhancement intensities at the crystal Bragg positions, the major change is primarily confined at the (200) position only. It can be rather easily visualised that in case of RuMnGa, where Mn is equally distributed in 4$a$ and 4$b$ with 50\% population at each sites. Since the magnetic intensities would be proportional to the magnetic structure factor (SF), the SF for (111) and (200) Bragg peaks in RuMnGa could be written as

M(4$a$) $-$ M(4$b$) for (111) magnetic Bragg peak

M(4$a$) + M(4$b$) for (200) magnetic Bragg peak

The above relations suggest that for a FiM spin arrangement, one should expect enhancement of intensity at the (111) Bragg reflection position, while the (200) peak would get enhanced for FM type of ordering.

Fig~\ref{fig:Neutron Fit} presents the ND spectra of RuMnGa taken at 300 K and 1.5 K. We see that while the intensity of (200) peak has been enhanced considerably, the (111) peak remain relatively unchanged. The powder diffraction data at low temperatures were thus analyzed by using both of the nuclear and magnetic phases, where the scale factor, lattice parameter and magnetic moment were refined along with overall thermal parameters. Our analysis suggest that the magnetic moment can be found only at the Mn sites with a moment value of 0.2(1) $\mu_{\rm B}$/f.u. along (100) direction, which matches well with the dc magnetisation data. Our analysis thus not only confirms the FM type of ordering, it also negates the possibility of magnetic moments on Ru and Ga atoms, contradicting the theoretically predicted description~\cite{ma2017computational}. The ND spectra analysis even rules out any antiparallel arrangement of Mn-moments at 4$a$ and 4$b$ sites as well. The slight unequal distribution of Mn atoms at 4$a$ and 4$b$ sites and the resultant slight deviation of VEC from 18 is likely to be the source of magnetic moment present in this system.

One may however notice that the ND spectra do not exhibit any noticeable change across $T_p$ and $T_s$, ostensibly due to a very small change in magnetism in a system that also have rather very low moment. However, the  signature in dc magnetic susceptibility indicate a disruption in spin arrangement, particularly at $T_p$, which could be a development of glassy phase or small spin-canting beyond the detection level of ND measurement technique. As ac susceptibility technique is considered to be very sensitive tool to detect such change, particularly for glassy phase, we have carried out this measurement  (see Sec.~\ref{sec:AC susceptibility}).

\subsection {Nonequilibrium Dynamics}\label{sec:NED}
\subsubsection{Magnetic Viscosity}

\begin{figure}[h]
\centering
\includegraphics [width=0.5\textwidth]{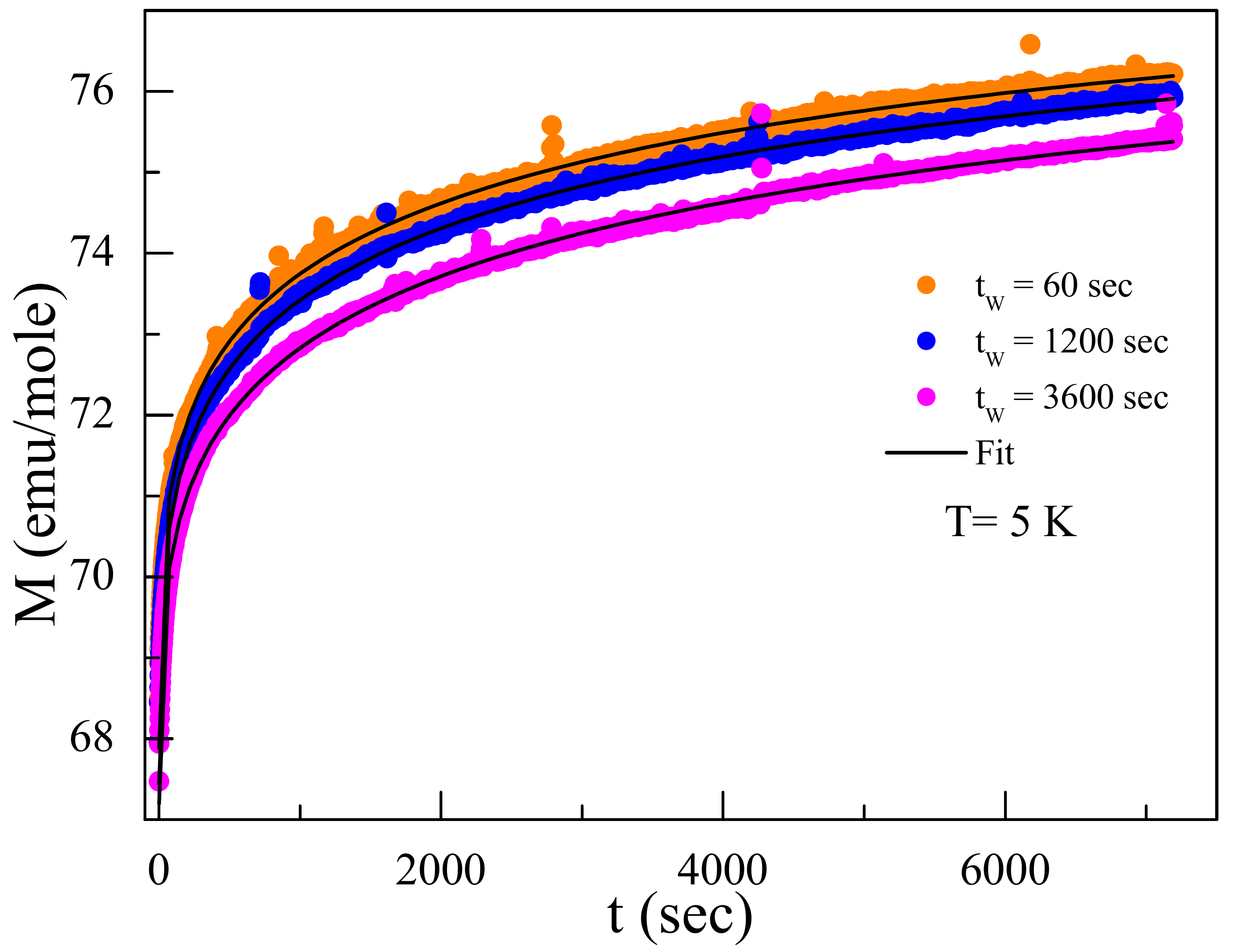}
\caption{Magnetic relaxation measured in Zero field cooled protocol (ZFC) at temperature T = 5 K for the wait time $t_W$ of 60 sec, 1200 sec and 3600 sec, respectively. The black line represents fit to the curve using KWW equation.}
\label{fig:Relaxation2}
\end{figure}

The temperature dependence of magnetic susceptibility, presented in Sec.~\ref{sec:DC magnetization}, depicts a small reduction of susceptibility below a temperature $T_p$ indicating a reduction in magnetic moment below this temperature. From the anomaly in $M(T)$ behavior at $T_p$, we had argued it to have come from either FiM type ordering, or spin reorientation or development of magnetic glassy phase. Since the ND measurements clearly ruled out the presence of FiM type of magnetic order, we have carried out magnetic relaxation measurements to check for the possible existence of magnetic glassy phase in RuMnGa. In this work, we have carried out the relaxation measurements in ZFC protocol, where the sample is cooled without applying  magnetic field from the paramagnetic region to below freezing temperature. After waiting for different time  intervals specified below at that temperature, a small amount of magnetic field is switched on, and the time evolution of magnetic moment, $M(t)$, is measured.  Fig.~\ref{fig:Relaxation2} show the relaxation behavior  at 5 K for different wait times $t_w$ = 60 s, 1200 s, and 3600 s. The presence of magnetic relaxation behavior clearly indicates the presence of magnetic glassy component in the system below T$_p^{ZFC}$. The magnetic relaxation data is analysed using standard Kohlrausch Williams-Watts (KWW) equation~\cite{pakhira2016large,mydosh1993spin,pakhira2020ferromagnetic}

\begin{equation}
M(t) = M_{0} \pm M_{g}\exp \left[-\left(\frac{t}{\tau}\right)^{\beta}\right]\label{eqn:relax}
\end{equation}

\begin{figure}[ht]
\centering
\includegraphics [width=0.5\textwidth]{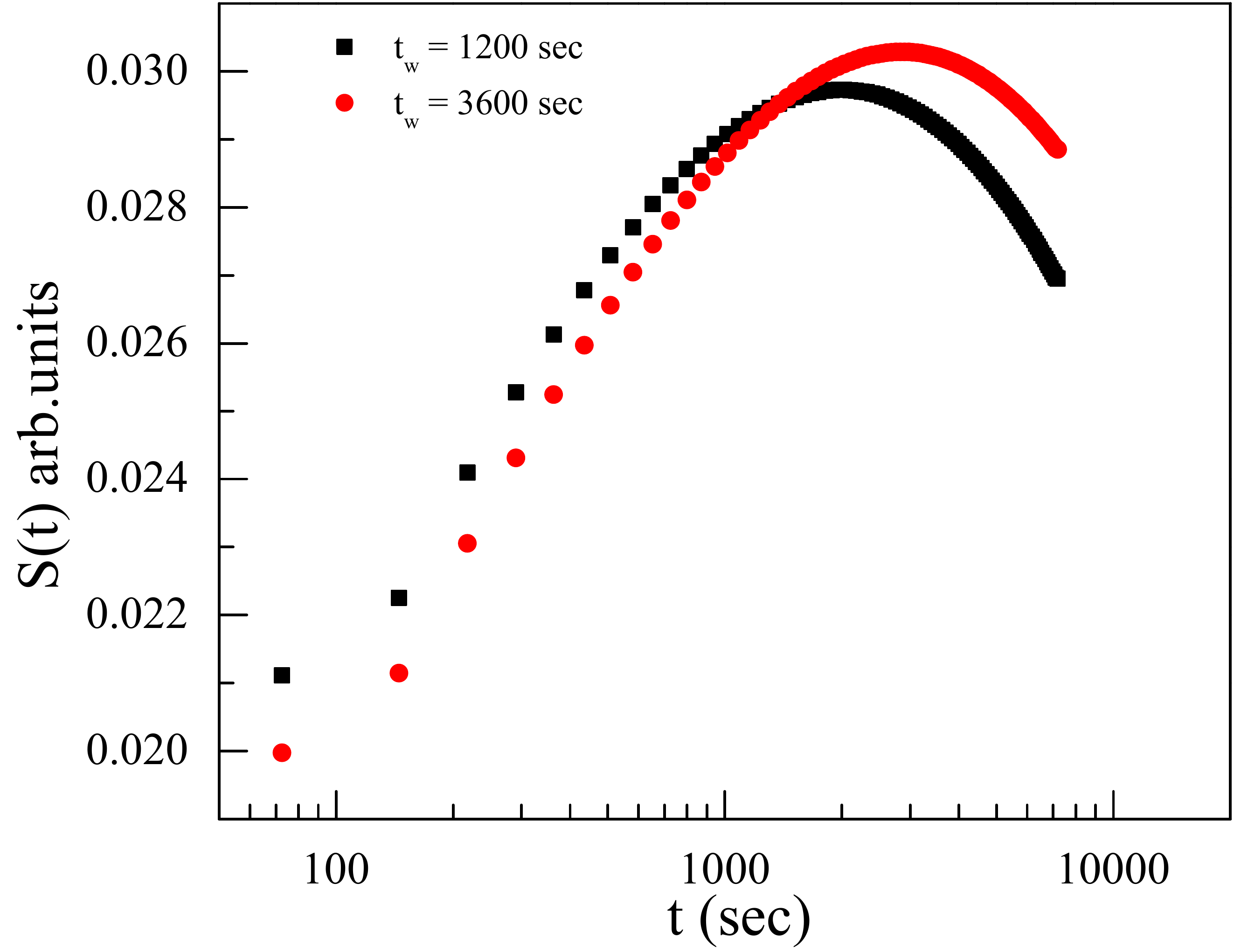}
\caption{Relaxation rate $S(t)$ calculated at T = 5 K for wait time 1200 sec and 3600 sec. Measured in 100 Oe in ZFC protocol.}
\label{fig:Viscosity}
\end{figure}
\noindent Here $M(t)$ is the magnetization data, $M_0$ is the intrinsic magnetic moment, $\tau$ is the relaxation time and $\beta$ is the stretch exponent. The fits to the experimental data are shown in black solid lines. The $\beta$ value at T = 5 K is found to be 0.32  and remains essentially unchanged for all the three measurements. The fractional value of $\beta$ (0 $<$ $\beta$ $<$ 1) suggest the glassy character of the system at 5 K~\cite{mydosh1993spin}. The $\tau$ value increases with the increase in wait time $t_w$ indicating that the system can remember the information about the wait time before the relaxation measurements begins. Such stiffening of spins i.e. the aging phenomenon implies that during its waiting time the system remains in a non-equilibrium dynamic state~\cite{pakhira2016large,gupta2023experimental}.

Fig.~\ref{fig:Viscosity} represents the magnetic viscosity curve, determined from the time evolution of magnetization for $t_w$ = 1200 s and 3600 s, show an inflection point at $t_w$. For measurements with larger wait time, this inflection point is also shifting towards higher observed time with larger wait time. This peak in magnetic viscosity curve is defined as $S(t)$ = $\frac{1}{H}$ $dM(t)/d(logt)$. This type of aging phenomenon confirms domain growth with time in RuMnGa~\cite{pakhira2016large,mydosh1993spin,MilyPRB,pakhira2018chemical,pakhira2020ferromagnetic}.

\subsubsection {Magnetic Memory}
\begin{figure}[ht]
\centering
\includegraphics [width=0.5\textwidth]{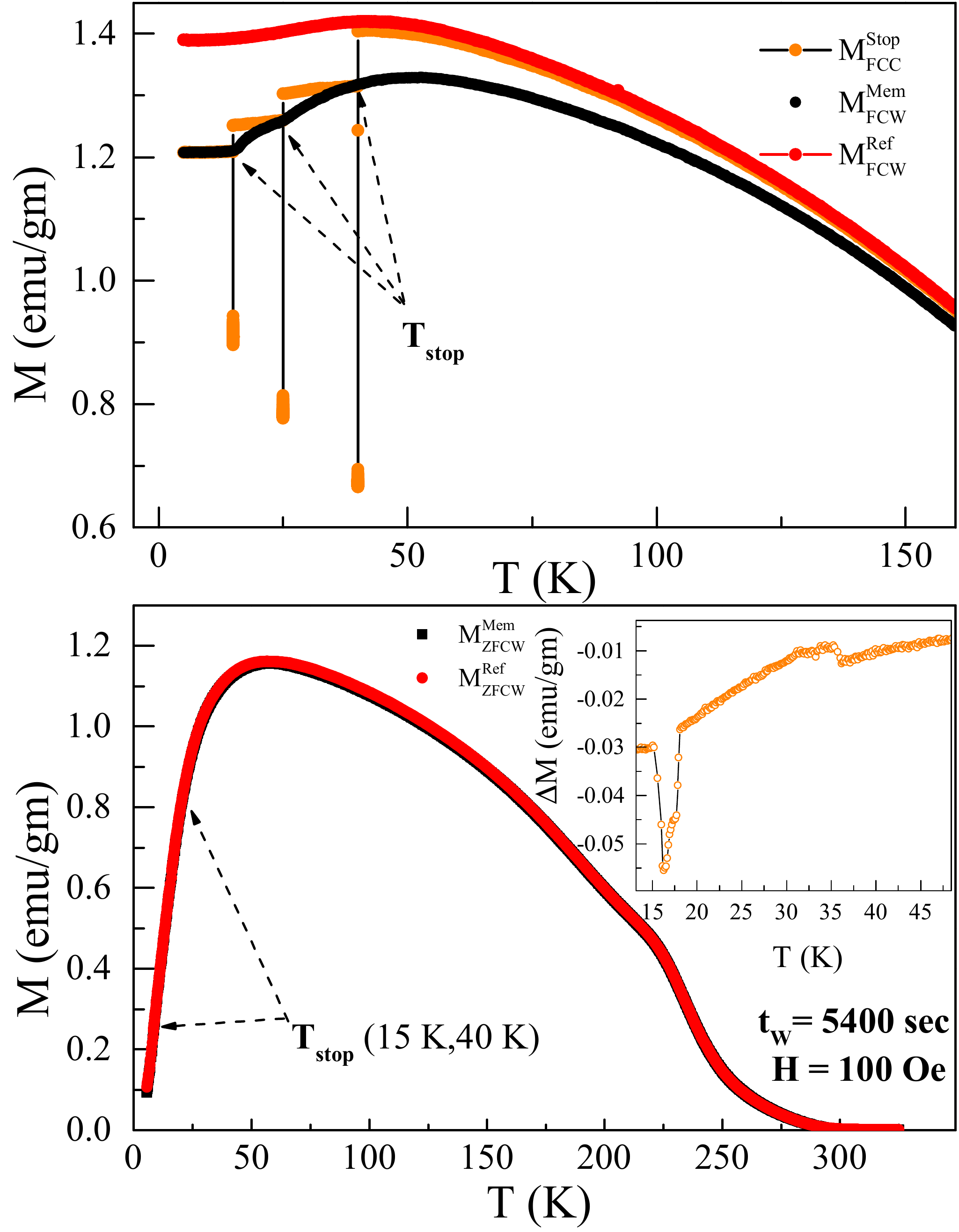}
\caption{(Upper panel) Magnetic memory effect measure in FC protocol using 100 Oe magnetic field with a stop time of 5400 sec at T = 15 K, 25 K and 40 K. (Lower panel) magnetic memory effect in ZFC protocol. Inset represents the difference between M$_{ZFCW}^{ref} (T)$ $-$ M$_{ZFCW}^{Mem} (T)$}
\label{fig:FCmem}
\end{figure}
To further examine the scenery of the magnetic state, we have performed ZFC and FC memory measurements. In FC memory experiment we have cooled down the sample from paramagnetic state (300 K) to 5 K in the presence of 100 Oe magnetic field. We have given in-between stop-times at T = 40 K, 25 K and 15 K for $t_w$= 5400 sec and the magnetization was measured. During $T_{stop}$ the magnetic field was switched off to zero and after the waiting time the same amount of magnetic field was reapplied. After reaching 5 K the sample was heated to the paramagnetic region with the same magnetic field (100 Oe), and the magnetization [$M(T)$] was measured again. This obtained curve reveal a clear signature of the magnetization history, as it shows anomaly at the temperatures where stops were applied during cooling. A reference curve is also shown as M$_{FCW}^{ref} (T)$. This indicated FC memory is present in RuMnGa in all the three $T_{stop}$ positions, viz., 15 K, 25 K and  40 K. This indicates a clear time-dependent magnetization phenomenon present in the studied system~\cite{pakhira2016large,pakhira2020ferromagnetic}.

Both the superparamagnetic as well as spin/cluster glass phases are known to exhibit such time-dependent magnetization behaviour in FC memory measurements. However, magnetic memory measurements under ZFC protocol can clearly distinguish between these two possibilities, as the memory effect would be absent in superparamagnetic material, but would manifest in spin/cluster glass systems~\cite{pakhira2016large,pakhira2020ferromagnetic}. In the ZFC memory measurement, the sample was cooled from paramagnetic region to 5K without applying any magnetic field and the intermediate stop times for 5400 sec was applied at $T_{stop}$ = 40 K and 15 K. After reaching 5 K, 100 Oe magnetic field was applied and sample was heated to paramagnetic state and magnetization was recorded. Here also magnetic anomaly is detected at 15 K and 40 K. For reference, the standard ZFC magnetization data M$_{ZFCW}^{ref} (T)$ was also recorded. Inset of fig.~\ref{fig:FCmem} (lower panel) shows $\triangle M$ (M$_{ZFCW}^{ref} (T)$ $-$ M$_{ZFCW}^{Mem} (T)$) indicating the memory effect to be present even in ZFC configuration and thus confirms spin-glass behavior in RuMnGa~\cite{pakhira2018chemical}.

\subsubsection {Magnetic Relaxation}
\label{sec:Rel}

\begin{figure}[ht]
\centering
\includegraphics [width=0.5\textwidth]{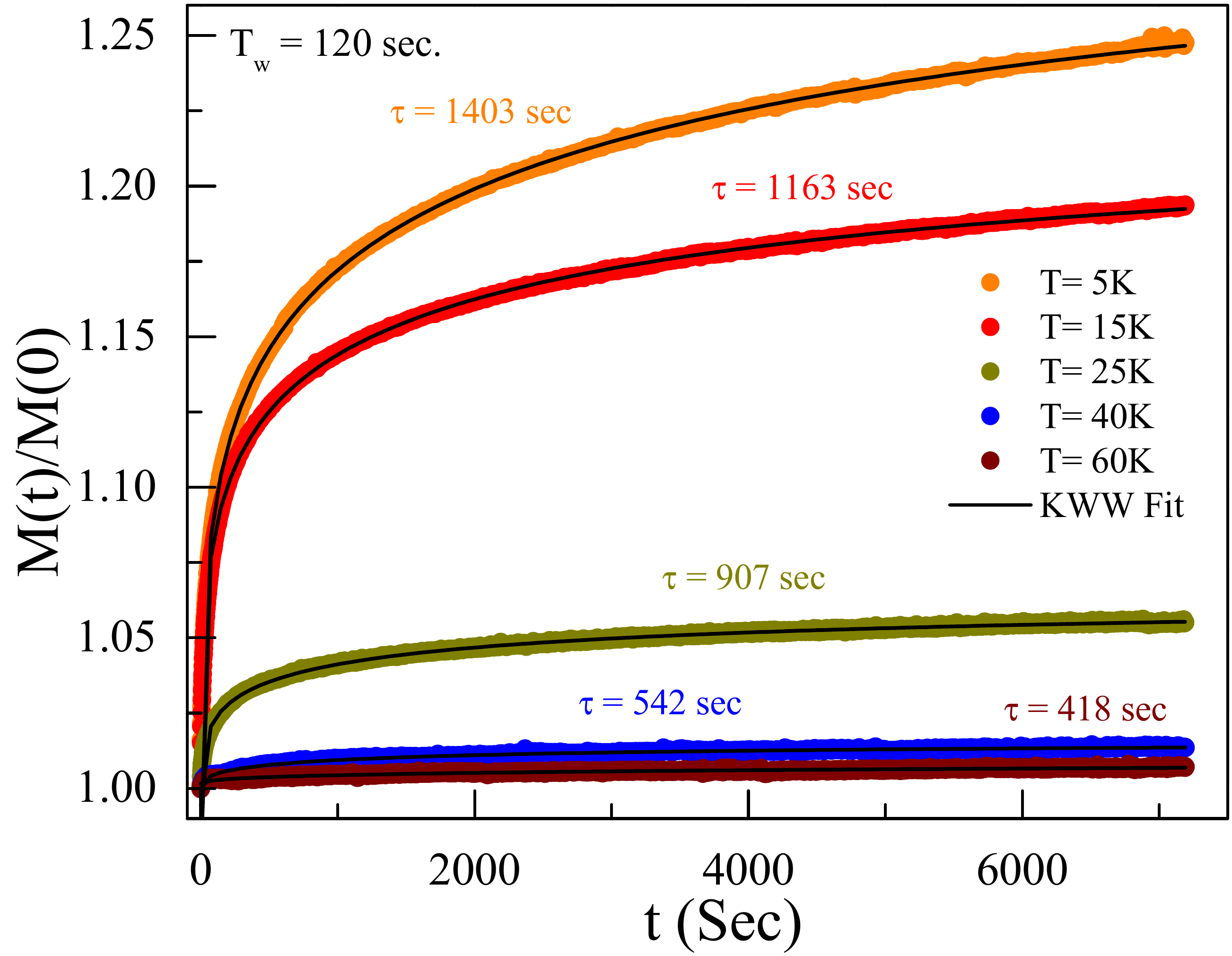}
\caption{Magnetic relaxation at ZFC protocol at various temperatures.}
\label{fig:Relaxation}
\end{figure}

Spin/cluster glass systems also show a prominent relaxation below $T_f$. Accordingly, we have studied magnetic relaxation behavior at different temperatures (T = 5 K, 15 K, 25 K, 40 K, and 60 K) in the ZFC protocol (Fig.~\ref{fig:Relaxation}) for $t_w$= 120 sec.  The magnetization value increases by nearly 25\%, 18\%, 5\% and 2\% at 5 K, 15 K, 25 K, and 40 K, respectively, after 7200 sec. these values are much larger than that reported earlier in another half-Heusler alloy, IrMnGa, in similar temperature region~\cite{kroder2019spin}.  We have fitted this experimental data using Eq.~\ref{eqn:relax}.

The value of $\beta$ value in the temperature range 5 - 60 found to change gradually between 0.32 to 0.38. The relaxation time decreases with the increase in measuring  temperature as expected in spin/cluster glass system~\cite{kroder2019spin}. Where, it may be noted that the value of $\beta$ can help us in distinguishing between the spin-glass and cluster-glass systems. In cannonical spin glass lies in the range 0.2-0.5~\cite{kroder2019spin}, and similar values are also estimated for RuMnGa. The value of $\tau$ of our sample also falls within the range reported earlier in another spin-glass system, CuMn~\cite{mydosh1993spin}.

\subsection {ac Susceptibility Study}
\label{sec:AC susceptibility}
\begin{figure}[h]
\centering
\includegraphics [width=0.5\textwidth]{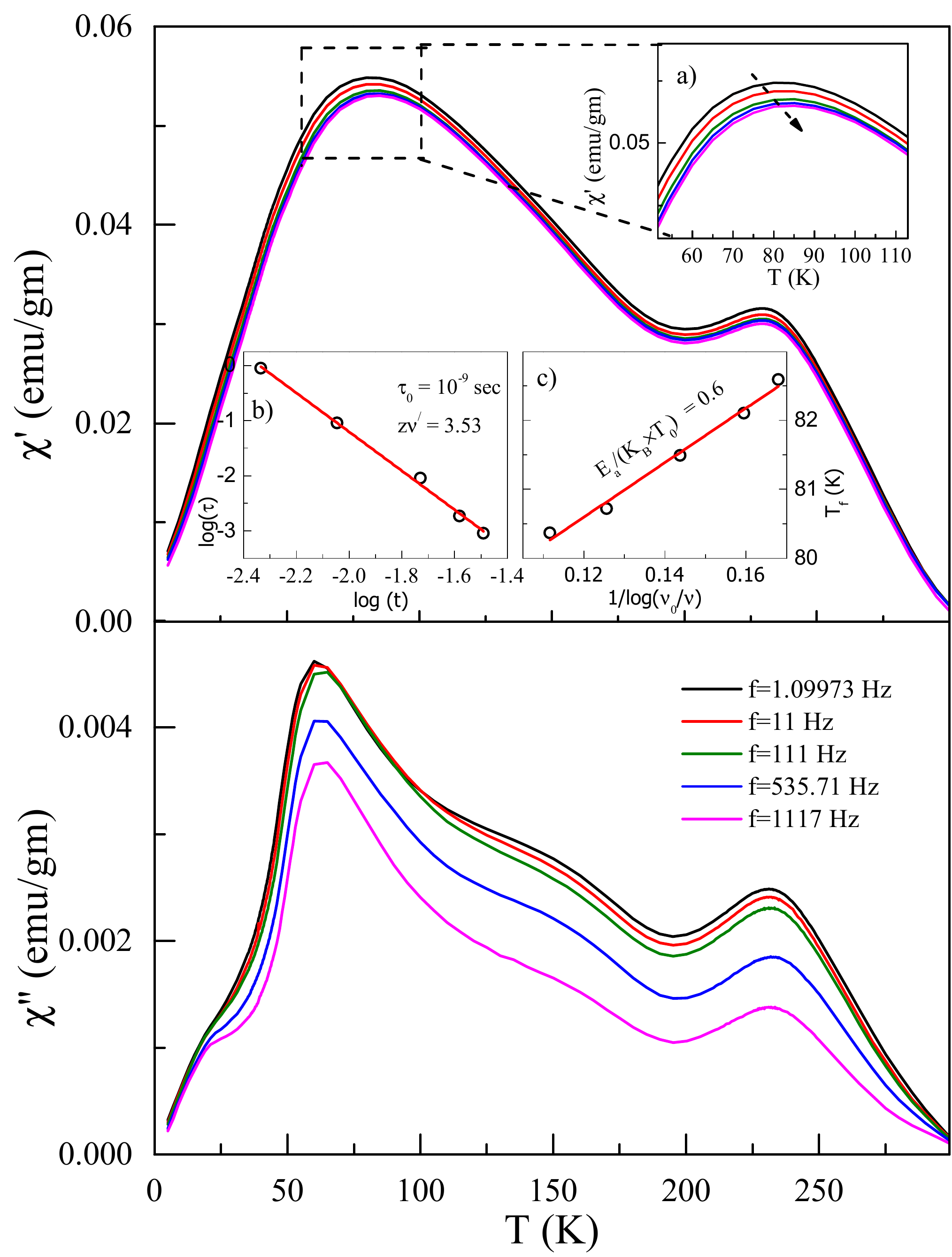}
\caption{(Top) the temperature variation of the real part of ac magnetic susceptibility data and (bottom) the imaginary part (loss component). Inset (a) shows the expanded view of the real part of low temperature AC susceptibility. The inset (b) depicts the plot of $\log(\tau)$ \textit{vs}. $\log(t)$ and inset (c) is the graphical representation of the Vogel-Fulcher law. The solid lines in both insets (b) and (c) are the linear fits of the two curves.}
\label{fig:AC}
\end{figure}
The ac susceptibility measurements are performed in the temperature region 5--300~K at different frequencies under an excitation field of 6 Oe (Fig.~\ref{fig:AC}). The $\chi^{\prime} (T)$ exhibits two peaks at $\sim$230~K and $\sim$80 K. The peak at high-temperature is observed at the same temperature $T_{\rm C}$, where the FM-like transition was observed in dc-magnetization data. Its temperature does not show any frequency dependence, as expected for a long range magnetic ordering. However, the low-temperature peak in the $\chi^{\prime} (T)$ observed at $T_f = 80$~K shifts to a higher temperature with increasing frequency, a typical signature of glassy transition~\cite{pakhira2016large}. The presence of glassy phase is also reflected in the dc magnetic measurement, as the peak in ZFC susceptibility (Fig.~\ref{fig:MT})  matches well with the peak of $\chi^{\prime \prime} (T)$ data. While the peak at $T_f$ indicates freezing of the ferromagnetic clusters related to the glassy phase, the peak in the $\chi^{\prime \prime} (T)$ at $T_{\rm C}$ is in accordance with the dissipative losses associated with the magnetic domain formation of the compound. Although no peak associated to $T_{s}$ is observed in the $\chi^{\prime} (T)$ data, a clear bend-like anomaly is observed in the corresponding $\chi^{\prime \prime} (T)$ behavior. Here we have to mention that also the temperature dependence of magnetic coercivity vanishes $T_{s}$ (Fig.~\ref{fig:MHFull}). Interestingly, the frequency dependence bifurcation of the magnitude of $\chi^{\prime \prime} (T)$ gets diminished below $T_{s}$. Thus, the ac-susceptibility data suggests the occurrence of two successive glass-like transitions below the long-ranged ordering temperature. Only a handful of systems are reported to exhibit such double glass-like transition~\cite{kumar2018evidence,murayama1986two,wang2006multiple}. The lack of anomaly in $\chi^{\prime} (T)$ data, and its presence in dc means that we need magnetic field to trigger it, indicating the anomaly at $T_{s}$ can arise due to the local spin reorientation in the system yielding a net ferromagnetic component.

To understand the nature of the spin-glass phase at $T_f$, the relative shift in freezing temperature (80 K at 1 Hz to 82.3 K at 1117 Hz) per decade of frequency is quantified by the Mydosh parameter, defined as~\cite{pakhira2016large,mydosh1993spin},

\begin{eqnarray}
\delta T_{f} =\frac{\Delta T_{f}}{T_{f}\Delta (\log_{10} f)}
\end{eqnarray}
where \textit{f} is the frequency applied and $T_{f}$ is the freezing temperature. The estimated value of $\delta T_{f}$ = 0.009 is quite small. Generally in cluster glass or superparamagnetic systems $\delta T_{f}$ value lies in between 0.1 to 0.3~\cite{pakhira2016large,pakhira2019spatially,chakraborty2022ground}, while a lower value suggest the canonical spin glass feature. For example, the $\delta T_{f}$ value for the well studied cannonical spin glass system CuMn is reported as 0.005~\cite{mydosh1993spin}, which is close to the value we have obtained for our studied compound.
To further ensure the cannonical spin-glass state formation in RuMnGa, dynamical scalling theory involving relaxation time ($\tau$) at any applied frequency (\textit{f}) and the spin-spin correlation length $\xi$ is utilised. As per this law~\cite{samanta2018reentrant,mondal2019physical}

\begin{eqnarray}
\tau = \tau_0\left(\frac{T_f-T_{SG}}{T_{SG}}\right)^{-z\nu^{\prime}}\label{eqn:dynamical scaling}
\end{eqnarray}

where $\tau$ = 1/\textit{f} and $\tau_0$ is the relaxation time corresponding to a single spin flip. $z\nu^{\prime}$ is the critical exponent for a correlation length $\xi = (\frac{T_{f}}{T_{SG}}-1)^{-\nu^{\prime}}$. In case of RuMnGa, $z\nu^{\prime}$ is found to be 3.62 and $\tau_0 \sim 10^{-9}$~s from the fitting shown in the inset (b) of Fig.~\ref{fig:AC}. The values are well within the range to those reported in different canonical spin-glass systems~\cite{mydosh1993spin,mondal2019physical,kroder2019spin}.

This frequency dependence of the spin-freezing temperature is also analyzed using Vogel-Fultcher law where the frequency variation is defined as

\begin{eqnarray}
f = f_{0} \exp \left[-\frac{E_a}{k_B(T_f-T_0)}\right]\label{eqn:Vogel-Fulcher}
\end{eqnarray}
$E_{a}$ is the activation energy and $T_{0}$ is the Vogel Fultcher temperature, $f_{0}$ is the characteriatic frequency, and k$_{\rm B}$ is the Boltzmann constant. As shown in inset (c) of Fig.~\ref{fig:AC}, the best fit is obtained with $E_a$/$k_{\rm B} = 39.49$~K and $T_0 = 75.85$~K, yielding $\frac{E_a}{k_{B}T_{0}} = 0.6$. For a cannonoical spin glassy system, one should expect this value to be less than 1~\cite{pakhira2016large,mondal2019physical}. So from Mydosh criteria, power law fitting and Vogel Fultcher it is found that the low temperature glassy transition around 80 K for RuMnGa is a cannonical spin-glass state formation.

\subsection {Neutron Depolarization}
\label{sec:NDP}
\begin{figure}[h]
\centering
\includegraphics [width=0.5\textwidth]{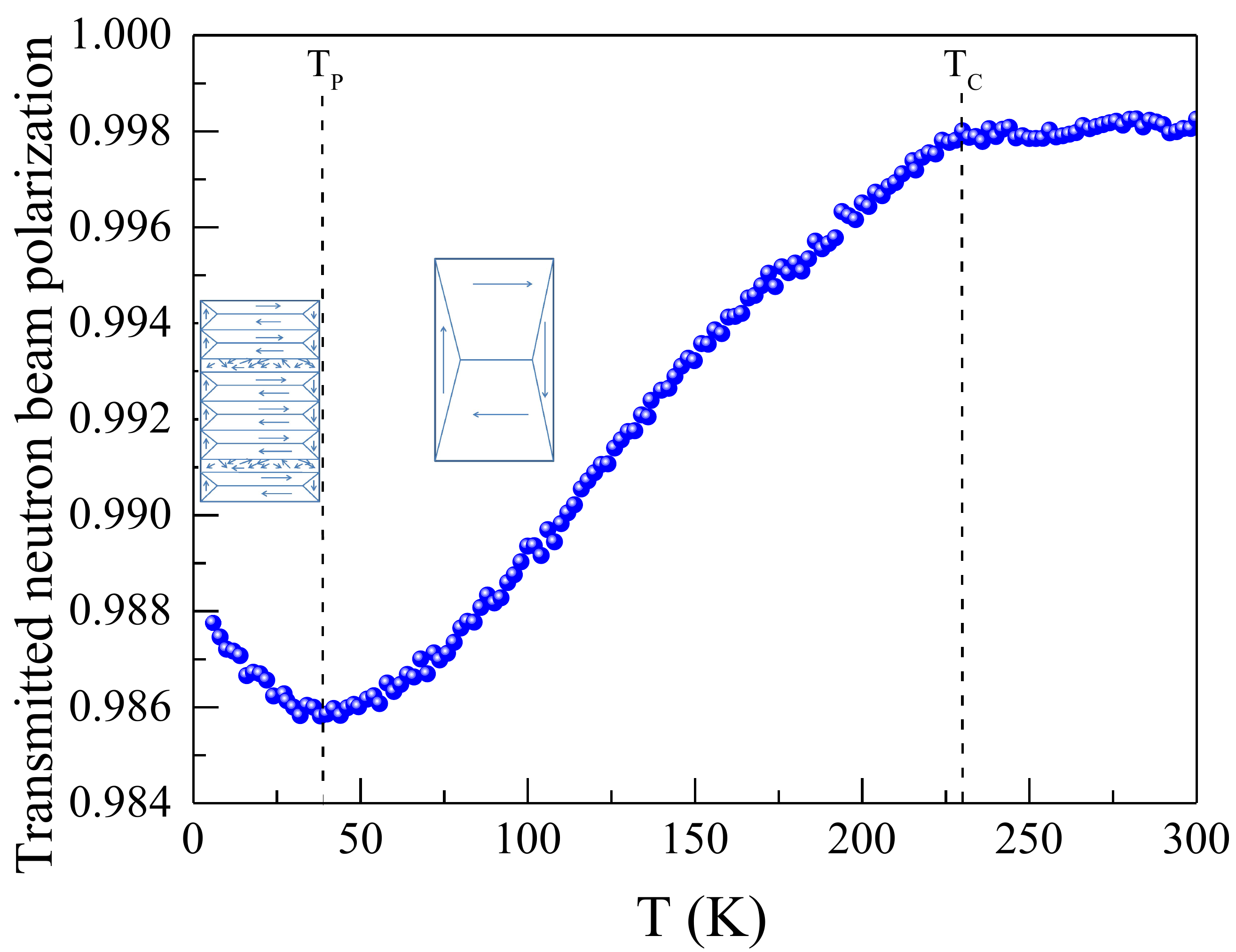}
\caption{Neutron depolarization curve taken between 5 - 300 K at 50 Oe applied magnetic field. Schematic diagrams of proposed domain structures for T$_P <$ T $<$ T$_C$ and T $<$ T$_P$ are also shown (see text).}
\label{fig:depo}
\end{figure}

The presence of magnetically glassy phase below $T_p$ poses an interesting conundrum whether the glassy phase originates due to a reentrant transition to glassy phase from an already ordered spin structure or the system exhibits magnetic inhomogeneity right from $T_{\rm C}$ due to the presence of disorder/vacancy in the system. In the later case, one may consider a fraction of the system exhibits long range ordering at $T_{\rm C}$, while a remaining fraction  undergoes the glassy transition at $T_p$. Neutron depolarisation measurement is a great tool to get insights into such complex magnetic state.  In this experiment the polarization vector in the neutron beam is studied after it transits across a magnetic medium. The magnetic inhomogeneities affects the polarization vector during its time of flight as the magnetic medium influences a net rotation of the polarization vector. When a spontaneous magnetization is present in a system, the value of polarization vector decreases which is known as depolarization~\cite{das2003neutron}. It is a perfect tool to distinguish between  FM/FiM and a glassy magnetic system. Magnetic moment fluctuations present in a paramagnetic state are too fast for neutron polarization as it can't keep up with the dissimilarity in the magnetic field acting on the neutron beam, so paramagnets do not show depolarization~\cite{das2003neutron,sk1999neutron}. In a glassy system too, as the magnetic spins are frozen randomly, the spacial fluctuations does not change the polarization vector of the neutron beam. In general, one can't expect depolarization in an AFM as there is no effective magnetic moment, whereas for a FM system the neutrons get depolarized while travelling through the randomly distributed domains~\cite{halpern1941passage,mitsuda1992neutron,samanta2018reentrant}. As we have already established RuMnGa to order ferromagnetically, any change in depolarisation at $T_p$ would confirm the reentrant character, where the glassy phase is formed at the expense of domain size reduction. No such change in depolarisation however could be sensed, if the ferromagnetic spin arrangement is not disturbed. During the measurement process, the sample was cooled under an applied field of 50 Oe and the depolarization data was recorded while warming the sample from 5 to 300 K. From the Fig.~\ref{fig:depo} it is clear that the polarization value starts to decrease from 250 K on cooling, signifying the gradual evolution of shapes and sizes of the magnetic domains in the system, as expected in a typical ferromagnetic system. The rather low extent of depolarization ($<$ 2\%), could be explained due to very weak moment in the system. The beam depolarization is related domain size and domain magnetization by the following relation:

\begin{eqnarray}
P_f = P_i \exp \left[-\alpha\left(\frac{d}{\delta}\right)<\phi_{\delta}>^2\right]
\end{eqnarray}

\noindent Here, $P_i$ = incident beam polarization, $\alpha$ = 1/3, $d$ = effective thickness of the sample in neutron beam, $\delta$ = average domain size, $\phi_{\delta}$ = (4.63 $\times$ 10$^{-10}$ G$^{-1}$ \AA$^{-2}$).$\lambda$.$B$.$\triangle$ = precession angle from single domain, $\lambda$ = 1.201 \AA = neutron wavelength, and $B$ = average domain magnetization (in Gauss) = 4$\pi$$M_S$ where $M_S$ = spontaneous magnetization of the sample in emu/cc = saturation magnetization (emu/g)$\times$ density (g/cc). The $\sim$ 2\% depolarization at 50 K for $d$ = 7 mm and $M_S$ = 25.4 emu/cc (from $M(H)$ curve at 50 K, Fig~\ref{fig:MHFull}) corresponds to an average one-dimensional magnetic domain size of 980 nm. Despite of such small depolarization effect, the polarization starts to recover close to $T_p$, suggesting the reduction of domain size below this temperature. In this process, there remain a finite probability that some of these newly formed domains can have sizes smaller than even the domain wall length, and hence can not satisfy the condition of proper domain formation. As a result, those magnetic spins tend to exhibit the glassy feature as found in the ac susceptibility measurements. As the ND measurement does not suggest any possible spin canting or growth of AFM phases at low temperature, one may consider the development of a reentrant spin-glass phase below this temperature in this compound~\cite{mitsuda1992neutron,samanta2018reentrant}. The signature of glassy phase at $T_p$ appear to overwhelm the weak signature of another glassy phase transition at $T_s$, as detected from ac susceptibility measurement (Sec.~\ref{sec:AC susceptibility}).

\subsection {Electrical Resistivity Study}
\label{sec:Resistivity}

\begin{figure}[h]
\centering
\includegraphics [width=0.5\textwidth]{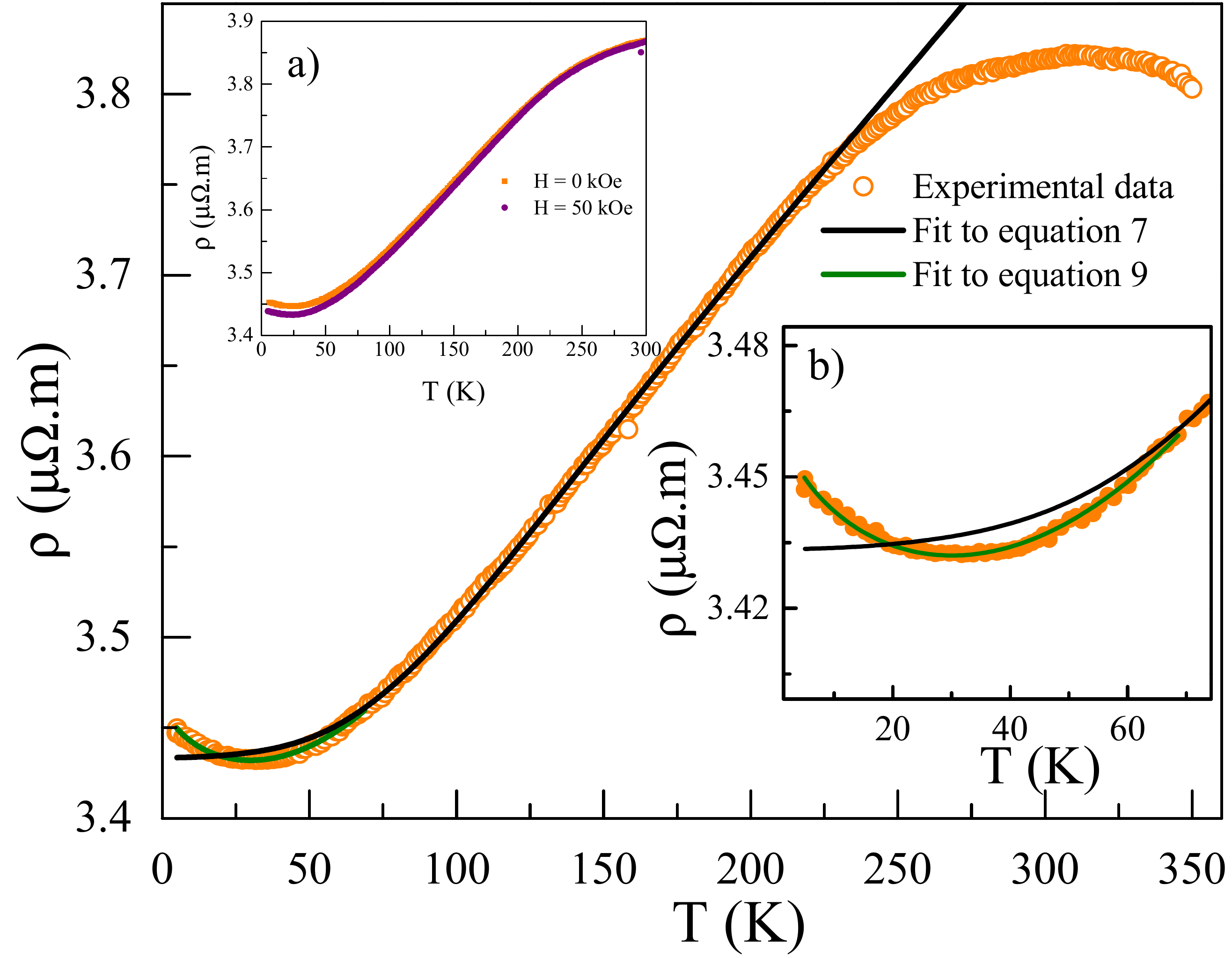}
\caption{Fitting of the zero field electrical resistivity  vs. temperature data measured in the temperature range from 5 K to 350 K. Inset a) represents the comparison of the resistivity as a function of temperature in the temperature region 5 - 300 K, obtained in 0 and 50 kOe magnetic fields. Inset b) depicts the low temperature region fits using two different equations.}
\label{fig:RT}
\end{figure}
Many of the FM Heusler alloys are also known to exhibit HMF properties, which can be experimentally sensed from their temperature dependence of resistivity, exhibiting minimal magnon contribution. The temperature dependence of zero-field resistivity behavior $\rho(T)$ measured in the range 5-300~K is shown in Fig.~\ref{fig:RT} in warming as well as cooling cycle. The $\rho$(T) behaviour does not exhibit any thermal hysteresis between these two measurement protocols. The residual resistivity ratio of R$_{5K}$/R$_{300K} = 1.28$ is consistent with the presence of significant structural disorder present in our system, as confirmed by XRD (Sec.~\ref{sec:structure}) and ND (Sec.~\ref{sec:ND}) analysis. In the paramagnetic region, above 250 K, $\rho(T)$ displays a very weak negative temperature coefficient of resistivity (TCR). As the system is cooled below its ferromagnetic ordering temperature, TCR changes its sign and become positive.

It may be noted here that only a handful of materials in HA family are known to exhibit such metal-semiconductor like transition across the magnetic ordering temperature~\cite{kawamiya1991electrical,nishino1993anomalous}. Even though a few theoretical model are proposed to elucidate this change in TCR coefficient by considering the magnetic spin scattering, none of them are universally acceptable~\cite{kataoka2001resistivity}. One of the most familiar theory that is applied to describe such change in TCR across T$_H$, as in the present case, is to consider the combined effect of spin dependent scattering, and electron-electron correlation~\cite{samanta2021itinerant}.

To describe the resistivity behaviour in the whole temperature range below its Curie temperature, the contribution of multiple scattering factors are generally considered, given by~\cite{rossiter1991electrical}
\begin{equation}
\rho(T) = \rho_0 + \rho_{ph}(T) +\rho_{mag}(T)
\label{eq1}
\end{equation}
\noindent
Here $\rho_{0}$ arises due to lattice defect contribution which is temperature independent, $\rho_{ph}$(T) is the scattering due to phonons and $\rho_{mag}$(T), is magnetic contribution, that originates from spin-flip transitions. According to Mattheissen rule, all these contribution to the electrical resistivity is additive in nature. $\rho_{ph}$ can be expressed as

\begin{equation}
\rho_{ph} = A{\bigg(\frac{T}{\Theta_{D}}\bigg)}^5 \int_{0}^{\frac{\Theta_{D}}{T}} \frac{x^{5}}{(e^{x}-1)(1-e^{-x})} dx
\label{RT}
\end{equation}
\noindent
where $A$ is a constant, and $\Theta_{\rm{D}}$ is the Debye temperature. $\rho_{mag}$(T) can be written as $BT^2$, where $B$ is a constant~\cite{bombor2013half}. The fit to Eq.~\ref{eq1} (Fig.~\ref{fig:RT}) confirms that the lattice contribution overwhelms magnon contribution. Although the low magnon contribution may tempt one to claim HMF character, one must keep it in mind that the Eq.~\ref{eq1} fits the data only in the high temperature region, 65 $<$ $T$ $<$ 240 K, whereas a proper HMF should have low magnon contribution even at the lowest temperature measured. The presence of shallow upturn around 40 K (Fig.~\ref{fig:RT}, inset (b)) thus negates the possibility of HMF in this compound.

The presence of such anomaly in $\rho$(T) at low temperature suggest the development of additional/different scattering mechanism below 40 K, resulting in the sign change of TCR again to positive. Similar behaviour has earlier  also been reported for a few  bulk and thin films of HA~\cite{bombor2013half,rani2018origin,slebarski2000weak,bainsla2014high} and generally explained by considering an additional presence of weak electron localization developed due to disorder induced coherent electron back scattering process that yields a T$^{\frac{1}{2}}$ dependence term in the resistivity~\cite{lee1985disordered}.
\begin{equation}
\rho(T) = \rho_0 + CT^2 -DT^{\frac{1}{2}}
\label{eq2}
\end{equation}
\noindent
The $\rho$(T) data is fitted in the region 5 $<$ $T$ $<$ 65 K using Eq.~\ref{eq2} and the results indicate that the major contribution in the resistivity for low temperature primarily arises due the weak localization phenomenon.

However, it may also be noted here that this low temperature upturn in $\rho$(T) coincides with the upturn in neutron depolarisation experiment. It is thus quite possible that the resistivity upturn may have its origin in the disruption of domain sizes below this temperature. The breaking of big domains into many smaller ones result in significant increase in domain wall area which could be responsible in enhanced scattering in electrical resistivity at the domain boundaries that acts as increased hindrance of electron flow. The application of magnetic field of 50 kOe do not affect the resistivity upturn (Fig.~\ref{fig:RT}, inset (a)) confirming the dominating role of magnetic phase boundaries over the formation of glassy phase.
\\

\section {\label{sec:Conclusion} Concluding Remarks}

In conclusion, we have synthesized half-Heusler alloy RuMnGa in single phase through arc melting technique. The combined studies of XRD and neutron diffraction reveal that the compound crystallizing in a disordered and non-stoichiometric composition as RuMn$_{0.86}$Ga$_{1.14}$. Despite the theoretically calculated positive formation energy reported earlier, the compound could only be synthesized experimentally with slightly higher lattice parameter than that predicted theoretically. Magnetic susceptibility, isothermal magnetization, neutron diffraction and neutron depolarization measurements reveal the compound exhibits a ferromagnetic to paramagnetic transition near 232 K and the spontaneous magnetic moment at 2 K is estimated to be 0.24 $\mu_{B}$/f.u. The time dependent dc magnetization indicates the presence of metastable spins at low temperature, whereas the ac magnetic susceptibility results confirms this to be canonical spin-glass phase. Neutron depolarization results indicate the spin-glass transition is reentrant in nature.  A weak ferromagnetic spin-reorientation phenomenon is also observed below $\sim$25 K, leading to a significant magnetic hysteresis at low temperatures. Interestingly, the electrical resistivity data indicate that the para- to ferromagnetic transition is also related with an insulator to metal-like transition whereas a weak localization is associated with the reentrant spin-glass transition temperature. Our Neutron diffraction measurement ruled out the FiM type of magnetic ordering proposed earlier by the theoretical study. The origins of ferromagnetic order and low saturation moment in this material have been explained to have originated from a minor off-stoichiometry in the composition.

\acknowledgments
Authors would like to pay gratitude to Late Prof. Vitalij K. Pecharsky, who participated in the initial stage of this research but could not finish due to his sad demise. The authors would also like to thank Prof. Subhradip Ghosh and Dr. Ashis Kundu of IIT Guwahati, India, for useful discussions. S.C. thanks University Grants Commission (UGC), Govt. Of India, for research fellowship. S.G. thanks SINP, Kolkata, for financial assistance. The research work conducted at Ames National Laboratory was supported by the U.S. Department of Energy, Office of Basic Energy Sciences, Division of Materials Sciences and Engineering. Ames National Laboratory is operated for the U.S. Department of Energy by Iowa State University under Contract No. DE-AC02-07CH11358.

\normalem
\bibliographystyle{apsrev4-2}

\begin{thebibliography}{53}%
\makeatletter
\providecommand \@ifxundefined [1]{%
 \@ifx{#1\undefined}
}%
\providecommand \@ifnum [1]{%
 \ifnum #1\expandafter \@firstoftwo
 \else \expandafter \@secondoftwo
 \fi
}%
\providecommand \@ifx [1]{%
 \ifx #1\expandafter \@firstoftwo
 \else \expandafter \@secondoftwo
 \fi
}%
\providecommand \natexlab [1]{#1}%
\providecommand \enquote  [1]{``#1''}%
\providecommand \bibnamefont  [1]{#1}%
\providecommand \bibfnamefont [1]{#1}%
\providecommand \citenamefont [1]{#1}%
\providecommand \href@noop [0]{\@secondoftwo}%
\providecommand \href [0]{\begingroup \@sanitize@url \@href}%
\providecommand \@href[1]{\@@startlink{#1}\@@href}%
\providecommand \@@href[1]{\endgroup#1\@@endlink}%
\providecommand \@sanitize@url [0]{\catcode `\\12\catcode `\$12\catcode
  `\&12\catcode `\#12\catcode `\^12\catcode `\_12\catcode `\%12\relax}%
\providecommand \@@startlink[1]{}%
\providecommand \@@endlink[0]{}%
\providecommand \url  [0]{\begingroup\@sanitize@url \@url }%
\providecommand \@url [1]{\endgroup\@href {#1}{\urlprefix }}%
\providecommand \urlprefix  [0]{URL }%
\providecommand \Eprint [0]{\href }%
\providecommand \doibase [0]{https://doi.org/}%
\providecommand \selectlanguage [0]{\@gobble}%
\providecommand \bibinfo  [0]{\@secondoftwo}%
\providecommand \bibfield  [0]{\@secondoftwo}%
\providecommand \translation [1]{[#1]}%
\providecommand \BibitemOpen [0]{}%
\providecommand \bibitemStop [0]{}%
\providecommand \bibitemNoStop [0]{.\EOS\space}%
\providecommand \EOS [0]{\spacefactor3000\relax}%
\providecommand \BibitemShut  [1]{\csname bibitem#1\endcsname}%
\let\auto@bib@innerbib\@empty
\bibitem [{\citenamefont {Katsnelson}\ \emph {et~al.}(2008)\citenamefont
  {Katsnelson}, \citenamefont {Irkhin}, \citenamefont {Chioncel}, \citenamefont
  {Lichtenstein},\ and\ \citenamefont {de~Groot}}]{katsnelson2008half}%
  \BibitemOpen
  \bibfield  {author} {\bibinfo {author} {\bibfnamefont {M.}~\bibnamefont
  {Katsnelson}}, \bibinfo {author} {\bibfnamefont {V.~Y.}\ \bibnamefont
  {Irkhin}}, \bibinfo {author} {\bibfnamefont {L.}~\bibnamefont {Chioncel}},
  \bibinfo {author} {\bibfnamefont {A.}~\bibnamefont {Lichtenstein}},\ and\
  \bibinfo {author} {\bibfnamefont {R.~A.}\ \bibnamefont {de~Groot}},\
  }\href@noop {} {\bibfield  {journal} {\bibinfo  {journal} {Reviews of Modern
  Physics}\ }\textbf {\bibinfo {volume} {80}},\ \bibinfo {pages} {315}
  (\bibinfo {year} {2008})}\BibitemShut {NoStop}%
\bibitem [{\citenamefont {De~Groot}\ \emph {et~al.}(1983)\citenamefont
  {De~Groot}, \citenamefont {Mueller}, \citenamefont {Van~Engen},\ and\
  \citenamefont {Buschow}}]{de1983new}%
  \BibitemOpen
  \bibfield  {author} {\bibinfo {author} {\bibfnamefont {R.}~\bibnamefont
  {De~Groot}}, \bibinfo {author} {\bibfnamefont {F.}~\bibnamefont {Mueller}},
  \bibinfo {author} {\bibfnamefont {P.}~\bibnamefont {Van~Engen}},\ and\
  \bibinfo {author} {\bibfnamefont {K.}~\bibnamefont {Buschow}},\ }\href@noop
  {} {\bibfield  {journal} {\bibinfo  {journal} {Physical Review Letters}\
  }\textbf {\bibinfo {volume} {50}},\ \bibinfo {pages} {2024} (\bibinfo {year}
  {1983})}\BibitemShut {NoStop}%
\bibitem [{\citenamefont {Mondal}\ \emph {et~al.}(2018)\citenamefont {Mondal},
  \citenamefont {Mazumdar}, \citenamefont {Ranganathan}, \citenamefont
  {Alleno}, \citenamefont {Sreeparvathy}, \citenamefont {Kanchana},\ and\
  \citenamefont {Vaitheeswaran}}]{mondal2018ferromagnetically}%
  \BibitemOpen
  \bibfield  {author} {\bibinfo {author} {\bibfnamefont {S.}~\bibnamefont
  {Mondal}}, \bibinfo {author} {\bibfnamefont {C.}~\bibnamefont {Mazumdar}},
  \bibinfo {author} {\bibfnamefont {R.}~\bibnamefont {Ranganathan}}, \bibinfo
  {author} {\bibfnamefont {E.}~\bibnamefont {Alleno}}, \bibinfo {author}
  {\bibfnamefont {P.}~\bibnamefont {Sreeparvathy}}, \bibinfo {author}
  {\bibfnamefont {V.}~\bibnamefont {Kanchana}},\ and\ \bibinfo {author}
  {\bibfnamefont {G.}~\bibnamefont {Vaitheeswaran}},\ }\href@noop {} {\bibfield
   {journal} {\bibinfo  {journal} {Physical Review B}\ }\textbf {\bibinfo
  {volume} {98}},\ \bibinfo {pages} {205130} (\bibinfo {year}
  {2018})}\BibitemShut {NoStop}%
\bibitem [{\citenamefont {Devi}\ \emph {et~al.}(2019)\citenamefont {Devi},
  \citenamefont {Mej{\'\i}a}, \citenamefont {Zavareh}, \citenamefont {Dubey},
  \citenamefont {Kushwaha}, \citenamefont {Skourski}, \citenamefont {Felser},
  \citenamefont {Nicklas},\ and\ \citenamefont {Singh}}]{devi2019improved}%
  \BibitemOpen
  \bibfield  {author} {\bibinfo {author} {\bibfnamefont {P.}~\bibnamefont
  {Devi}}, \bibinfo {author} {\bibfnamefont {C.~S.}\ \bibnamefont
  {Mej{\'\i}a}}, \bibinfo {author} {\bibfnamefont {M.~G.}\ \bibnamefont
  {Zavareh}}, \bibinfo {author} {\bibfnamefont {K.}~\bibnamefont {Dubey}},
  \bibinfo {author} {\bibfnamefont {P.}~\bibnamefont {Kushwaha}}, \bibinfo
  {author} {\bibfnamefont {Y.}~\bibnamefont {Skourski}}, \bibinfo {author}
  {\bibfnamefont {C.}~\bibnamefont {Felser}}, \bibinfo {author} {\bibfnamefont
  {M.}~\bibnamefont {Nicklas}},\ and\ \bibinfo {author} {\bibfnamefont
  {S.}~\bibnamefont {Singh}},\ }\href@noop {} {\bibfield  {journal} {\bibinfo
  {journal} {Physical Review Materials}\ }\textbf {\bibinfo {volume} {3}},\
  \bibinfo {pages} {062401} (\bibinfo {year} {2019})}\BibitemShut {NoStop}%
\bibitem [{\citenamefont {Aksoy}\ \emph {et~al.}(2009)\citenamefont {Aksoy},
  \citenamefont {Acet}, \citenamefont {Deen}, \citenamefont {Ma{\~n}osa},\ and\
  \citenamefont {Planes}}]{aksoy2009magnetic}%
  \BibitemOpen
  \bibfield  {author} {\bibinfo {author} {\bibfnamefont {S.}~\bibnamefont
  {Aksoy}}, \bibinfo {author} {\bibfnamefont {M.}~\bibnamefont {Acet}},
  \bibinfo {author} {\bibfnamefont {P.}~\bibnamefont {Deen}}, \bibinfo {author}
  {\bibfnamefont {L.}~\bibnamefont {Ma{\~n}osa}},\ and\ \bibinfo {author}
  {\bibfnamefont {A.}~\bibnamefont {Planes}},\ }\href@noop {} {\bibfield
  {journal} {\bibinfo  {journal} {Physical Review B}\ }\textbf {\bibinfo
  {volume} {79}},\ \bibinfo {pages} {212401} (\bibinfo {year}
  {2009})}\BibitemShut {NoStop}%
\bibitem [{\citenamefont {Planes}\ \emph {et~al.}(2009)\citenamefont {Planes},
  \citenamefont {Ma{\~n}osa},\ and\ \citenamefont
  {Acet}}]{planes2009magnetocaloric}%
  \BibitemOpen
  \bibfield  {author} {\bibinfo {author} {\bibfnamefont {A.}~\bibnamefont
  {Planes}}, \bibinfo {author} {\bibfnamefont {L.}~\bibnamefont {Ma{\~n}osa}},\
  and\ \bibinfo {author} {\bibfnamefont {M.}~\bibnamefont {Acet}},\ }\href@noop
  {} {\bibfield  {journal} {\bibinfo  {journal} {Journal of Physics: Condensed
  Matter}\ }\textbf {\bibinfo {volume} {21}},\ \bibinfo {pages} {233201}
  (\bibinfo {year} {2009})}\BibitemShut {NoStop}%
\bibitem [{\citenamefont {Wang}\ \emph {et~al.}(2016)\citenamefont {Wang},
  \citenamefont {Vergniory}, \citenamefont {Kushwaha}, \citenamefont
  {Hirschberger}, \citenamefont {Chulkov}, \citenamefont {Ernst}, \citenamefont
  {Ong}, \citenamefont {Cava},\ and\ \citenamefont {Bernevig}}]{wang2016time}%
  \BibitemOpen
  \bibfield  {author} {\bibinfo {author} {\bibfnamefont {Z.}~\bibnamefont
  {Wang}}, \bibinfo {author} {\bibfnamefont {M.}~\bibnamefont {Vergniory}},
  \bibinfo {author} {\bibfnamefont {S.}~\bibnamefont {Kushwaha}}, \bibinfo
  {author} {\bibfnamefont {M.}~\bibnamefont {Hirschberger}}, \bibinfo {author}
  {\bibfnamefont {E.}~\bibnamefont {Chulkov}}, \bibinfo {author} {\bibfnamefont
  {A.}~\bibnamefont {Ernst}}, \bibinfo {author} {\bibfnamefont {N.~P.}\
  \bibnamefont {Ong}}, \bibinfo {author} {\bibfnamefont {R.~J.}\ \bibnamefont
  {Cava}},\ and\ \bibinfo {author} {\bibfnamefont {B.~A.}\ \bibnamefont
  {Bernevig}},\ }\href@noop {} {\bibfield  {journal} {\bibinfo  {journal}
  {Physical Review Letters}\ }\textbf {\bibinfo {volume} {117}},\ \bibinfo
  {pages} {236401} (\bibinfo {year} {2016})}\BibitemShut {NoStop}%
\bibitem [{\citenamefont {Venkateswara}\ \emph {et~al.}(2019)\citenamefont
  {Venkateswara}, \citenamefont {Samatham}, \citenamefont {Babu}, \citenamefont
  {Suresh},\ and\ \citenamefont {Alam}}]{venkateswara2019coexistence}%
  \BibitemOpen
  \bibfield  {author} {\bibinfo {author} {\bibfnamefont {Y.}~\bibnamefont
  {Venkateswara}}, \bibinfo {author} {\bibfnamefont {S.~S.}\ \bibnamefont
  {Samatham}}, \bibinfo {author} {\bibfnamefont {P.}~\bibnamefont {Babu}},
  \bibinfo {author} {\bibfnamefont {K.}~\bibnamefont {Suresh}},\ and\ \bibinfo
  {author} {\bibfnamefont {A.}~\bibnamefont {Alam}},\ }\href@noop {} {\bibfield
   {journal} {\bibinfo  {journal} {Physical Review B}\ }\textbf {\bibinfo
  {volume} {100}},\ \bibinfo {pages} {180404} (\bibinfo {year}
  {2019})}\BibitemShut {NoStop}%
\bibitem [{\citenamefont {Meshcheriakova}\ \emph {et~al.}(2014)\citenamefont
  {Meshcheriakova}, \citenamefont {Chadov}, \citenamefont {Nayak},
  \citenamefont {R{\"o}{\ss}ler}, \citenamefont {K{\"u}bler}, \citenamefont
  {Andr{\'e}}, \citenamefont {Tsirlin}, \citenamefont {Kiss}, \citenamefont
  {Hausdorf}, \citenamefont {Kalache} \emph
  {et~al.}}]{meshcheriakova2014large}%
  \BibitemOpen
  \bibfield  {author} {\bibinfo {author} {\bibfnamefont {O.}~\bibnamefont
  {Meshcheriakova}}, \bibinfo {author} {\bibfnamefont {S.}~\bibnamefont
  {Chadov}}, \bibinfo {author} {\bibfnamefont {A.}~\bibnamefont {Nayak}},
  \bibinfo {author} {\bibfnamefont {U.}~\bibnamefont {R{\"o}{\ss}ler}},
  \bibinfo {author} {\bibfnamefont {J.}~\bibnamefont {K{\"u}bler}}, \bibinfo
  {author} {\bibfnamefont {G.}~\bibnamefont {Andr{\'e}}}, \bibinfo {author}
  {\bibfnamefont {A.}~\bibnamefont {Tsirlin}}, \bibinfo {author} {\bibfnamefont
  {J.}~\bibnamefont {Kiss}}, \bibinfo {author} {\bibfnamefont {S.}~\bibnamefont
  {Hausdorf}}, \bibinfo {author} {\bibfnamefont {A.}~\bibnamefont {Kalache}},
  \emph {et~al.},\ }\href@noop {} {\bibfield  {journal} {\bibinfo  {journal}
  {Physical Review Letters}\ }\textbf {\bibinfo {volume} {113}},\ \bibinfo
  {pages} {087203} (\bibinfo {year} {2014})}\BibitemShut {NoStop}%
\bibitem [{\citenamefont {Zuo}\ \emph {et~al.}(2018)\citenamefont {Zuo},
  \citenamefont {Liang}, \citenamefont {Zhang}, \citenamefont {Peng},
  \citenamefont {Xiong}, \citenamefont {Liu}, \citenamefont {Li}, \citenamefont
  {Zhao}, \citenamefont {Sun}, \citenamefont {Hu} \emph
  {et~al.}}]{zuo2018zero}%
  \BibitemOpen
  \bibfield  {author} {\bibinfo {author} {\bibfnamefont {S.}~\bibnamefont
  {Zuo}}, \bibinfo {author} {\bibfnamefont {F.}~\bibnamefont {Liang}}, \bibinfo
  {author} {\bibfnamefont {Y.}~\bibnamefont {Zhang}}, \bibinfo {author}
  {\bibfnamefont {L.}~\bibnamefont {Peng}}, \bibinfo {author} {\bibfnamefont
  {J.}~\bibnamefont {Xiong}}, \bibinfo {author} {\bibfnamefont
  {Y.}~\bibnamefont {Liu}}, \bibinfo {author} {\bibfnamefont {R.}~\bibnamefont
  {Li}}, \bibinfo {author} {\bibfnamefont {T.}~\bibnamefont {Zhao}}, \bibinfo
  {author} {\bibfnamefont {J.}~\bibnamefont {Sun}}, \bibinfo {author}
  {\bibfnamefont {F.}~\bibnamefont {Hu}}, \emph {et~al.},\ }\href@noop {}
  {\bibfield  {journal} {\bibinfo  {journal} {Physical Review Materials}\
  }\textbf {\bibinfo {volume} {2}},\ \bibinfo {pages} {104408} (\bibinfo {year}
  {2018})}\BibitemShut {NoStop}%
\bibitem [{\citenamefont {Gupta}\ \emph {et~al.}(2022)\citenamefont {Gupta},
  \citenamefont {Chakraborty}, \citenamefont {Pakhira}, \citenamefont
  {Barreteau}, \citenamefont {Crivello}, \citenamefont {Bandyopadhyay},
  \citenamefont {Greneche}, \citenamefont {Alleno},\ and\ \citenamefont
  {Mazumdar}}]{gupta2022coexisting}%
  \BibitemOpen
  \bibfield  {author} {\bibinfo {author} {\bibfnamefont {S.}~\bibnamefont
  {Gupta}}, \bibinfo {author} {\bibfnamefont {S.}~\bibnamefont {Chakraborty}},
  \bibinfo {author} {\bibfnamefont {S.}~\bibnamefont {Pakhira}}, \bibinfo
  {author} {\bibfnamefont {C.}~\bibnamefont {Barreteau}}, \bibinfo {author}
  {\bibfnamefont {J.-C.}\ \bibnamefont {Crivello}}, \bibinfo {author}
  {\bibfnamefont {B.}~\bibnamefont {Bandyopadhyay}}, \bibinfo {author}
  {\bibfnamefont {J.~M.}\ \bibnamefont {Greneche}}, \bibinfo {author}
  {\bibfnamefont {E.}~\bibnamefont {Alleno}},\ and\ \bibinfo {author}
  {\bibfnamefont {C.}~\bibnamefont {Mazumdar}},\ }\href@noop {} {\bibfield
  {journal} {\bibinfo  {journal} {Physical Review B}\ }\textbf {\bibinfo
  {volume} {106}},\ \bibinfo {pages} {115148} (\bibinfo {year}
  {2022})}\BibitemShut {NoStop}%
\bibitem [{\citenamefont {Graf}\ \emph {et~al.}(2011)\citenamefont {Graf},
  \citenamefont {Felser},\ and\ \citenamefont {Parkin}}]{graf2011simple}%
  \BibitemOpen
  \bibfield  {author} {\bibinfo {author} {\bibfnamefont {T.}~\bibnamefont
  {Graf}}, \bibinfo {author} {\bibfnamefont {C.}~\bibnamefont {Felser}},\ and\
  \bibinfo {author} {\bibfnamefont {S.~S.}\ \bibnamefont {Parkin}},\
  }\href@noop {} {\bibfield  {journal} {\bibinfo  {journal} {Progress in Solid
  State Chemistry}\ }\textbf {\bibinfo {volume} {39}},\ \bibinfo {pages} {1}
  (\bibinfo {year} {2011})}\BibitemShut {NoStop}%
\bibitem [{\citenamefont {Park}\ \emph {et~al.}(2011)\citenamefont {Park},
  \citenamefont {Wunderlich}, \citenamefont {Mart{\'\i}}, \citenamefont
  {Hol{\`y}}, \citenamefont {Kurosaki}, \citenamefont {Yamada}, \citenamefont
  {Yamamoto}, \citenamefont {Nishide}, \citenamefont {Hayakawa}, \citenamefont
  {Takahashi} \emph {et~al.}}]{park2011spin}%
  \BibitemOpen
  \bibfield  {author} {\bibinfo {author} {\bibfnamefont {B.~G.}\ \bibnamefont
  {Park}}, \bibinfo {author} {\bibfnamefont {J.}~\bibnamefont {Wunderlich}},
  \bibinfo {author} {\bibfnamefont {X.}~\bibnamefont {Mart{\'\i}}}, \bibinfo
  {author} {\bibfnamefont {V.}~\bibnamefont {Hol{\`y}}}, \bibinfo {author}
  {\bibfnamefont {Y.}~\bibnamefont {Kurosaki}}, \bibinfo {author}
  {\bibfnamefont {M.}~\bibnamefont {Yamada}}, \bibinfo {author} {\bibfnamefont
  {H.}~\bibnamefont {Yamamoto}}, \bibinfo {author} {\bibfnamefont
  {A.}~\bibnamefont {Nishide}}, \bibinfo {author} {\bibfnamefont
  {J.}~\bibnamefont {Hayakawa}}, \bibinfo {author} {\bibfnamefont
  {H.}~\bibnamefont {Takahashi}}, \emph {et~al.},\ }\href@noop {} {\bibfield
  {journal} {\bibinfo  {journal} {Nature Materials}\ }\textbf {\bibinfo
  {volume} {10}},\ \bibinfo {pages} {347} (\bibinfo {year} {2011})}\BibitemShut
  {NoStop}%
\bibitem [{\citenamefont {Baltz}\ \emph {et~al.}(2018)\citenamefont {Baltz},
  \citenamefont {Manchon}, \citenamefont {Tsoi}, \citenamefont {Moriyama},
  \citenamefont {Ono},\ and\ \citenamefont
  {Tserkovnyak}}]{baltz2018antiferromagnetic}%
  \BibitemOpen
  \bibfield  {author} {\bibinfo {author} {\bibfnamefont {V.}~\bibnamefont
  {Baltz}}, \bibinfo {author} {\bibfnamefont {A.}~\bibnamefont {Manchon}},
  \bibinfo {author} {\bibfnamefont {M.}~\bibnamefont {Tsoi}}, \bibinfo {author}
  {\bibfnamefont {T.}~\bibnamefont {Moriyama}}, \bibinfo {author}
  {\bibfnamefont {T.}~\bibnamefont {Ono}},\ and\ \bibinfo {author}
  {\bibfnamefont {Y.}~\bibnamefont {Tserkovnyak}},\ }\href@noop {} {\bibfield
  {journal} {\bibinfo  {journal} {Reviews of Modern Physics}\ }\textbf
  {\bibinfo {volume} {90}},\ \bibinfo {pages} {015005} (\bibinfo {year}
  {2018})}\BibitemShut {NoStop}%
\bibitem [{\citenamefont {Wang}\ \emph {et~al.}(2012)\citenamefont {Wang},
  \citenamefont {Song}, \citenamefont {Cui}, \citenamefont {Wang},
  \citenamefont {Zeng},\ and\ \citenamefont {Pan}}]{wang2012room}%
  \BibitemOpen
  \bibfield  {author} {\bibinfo {author} {\bibfnamefont {Y.}~\bibnamefont
  {Wang}}, \bibinfo {author} {\bibfnamefont {C.}~\bibnamefont {Song}}, \bibinfo
  {author} {\bibfnamefont {B.}~\bibnamefont {Cui}}, \bibinfo {author}
  {\bibfnamefont {G.}~\bibnamefont {Wang}}, \bibinfo {author} {\bibfnamefont
  {F.}~\bibnamefont {Zeng}},\ and\ \bibinfo {author} {\bibfnamefont
  {F.}~\bibnamefont {Pan}},\ }\href@noop {} {\bibfield  {journal} {\bibinfo
  {journal} {Physical Review Letters}\ }\textbf {\bibinfo {volume} {109}},\
  \bibinfo {pages} {137201} (\bibinfo {year} {2012})}\BibitemShut {NoStop}%
\bibitem [{\citenamefont {Cai}\ \emph {et~al.}(2020)\citenamefont {Cai},
  \citenamefont {Zhu}, \citenamefont {Lee}, \citenamefont {Mishra},
  \citenamefont {Ren}, \citenamefont {Pollard}, \citenamefont {He},
  \citenamefont {Liang}, \citenamefont {Teo},\ and\ \citenamefont
  {Yang}}]{cai2020ultrafast}%
  \BibitemOpen
  \bibfield  {author} {\bibinfo {author} {\bibfnamefont {K.}~\bibnamefont
  {Cai}}, \bibinfo {author} {\bibfnamefont {Z.}~\bibnamefont {Zhu}}, \bibinfo
  {author} {\bibfnamefont {J.~M.}\ \bibnamefont {Lee}}, \bibinfo {author}
  {\bibfnamefont {R.}~\bibnamefont {Mishra}}, \bibinfo {author} {\bibfnamefont
  {L.}~\bibnamefont {Ren}}, \bibinfo {author} {\bibfnamefont {S.~D.}\
  \bibnamefont {Pollard}}, \bibinfo {author} {\bibfnamefont {P.}~\bibnamefont
  {He}}, \bibinfo {author} {\bibfnamefont {G.}~\bibnamefont {Liang}}, \bibinfo
  {author} {\bibfnamefont {K.~L.}\ \bibnamefont {Teo}},\ and\ \bibinfo {author}
  {\bibfnamefont {H.}~\bibnamefont {Yang}},\ }\href@noop {} {\bibfield
  {journal} {\bibinfo  {journal} {Nature Electronics}\ }\textbf {\bibinfo
  {volume} {3}},\ \bibinfo {pages} {37} (\bibinfo {year} {2020})}\BibitemShut
  {NoStop}%
\bibitem [{\citenamefont {Nayak}\ \emph {et~al.}(2015)\citenamefont {Nayak},
  \citenamefont {Nicklas}, \citenamefont {Chadov}, \citenamefont {Khuntia},
  \citenamefont {Shekhar}, \citenamefont {Kalache}, \citenamefont {Baenitz},
  \citenamefont {Skourski}, \citenamefont {Guduru}, \citenamefont {Puri} \emph
  {et~al.}}]{nayak2015design}%
  \BibitemOpen
  \bibfield  {author} {\bibinfo {author} {\bibfnamefont {A.~K.}\ \bibnamefont
  {Nayak}}, \bibinfo {author} {\bibfnamefont {M.}~\bibnamefont {Nicklas}},
  \bibinfo {author} {\bibfnamefont {S.}~\bibnamefont {Chadov}}, \bibinfo
  {author} {\bibfnamefont {P.}~\bibnamefont {Khuntia}}, \bibinfo {author}
  {\bibfnamefont {C.}~\bibnamefont {Shekhar}}, \bibinfo {author} {\bibfnamefont
  {A.}~\bibnamefont {Kalache}}, \bibinfo {author} {\bibfnamefont
  {M.}~\bibnamefont {Baenitz}}, \bibinfo {author} {\bibfnamefont
  {Y.}~\bibnamefont {Skourski}}, \bibinfo {author} {\bibfnamefont {V.~K.}\
  \bibnamefont {Guduru}}, \bibinfo {author} {\bibfnamefont {A.}~\bibnamefont
  {Puri}}, \emph {et~al.},\ }\href@noop {} {\bibfield  {journal} {\bibinfo
  {journal} {Nature Materials}\ }\textbf {\bibinfo {volume} {14}},\ \bibinfo
  {pages} {679} (\bibinfo {year} {2015})}\BibitemShut {NoStop}%
\bibitem [{\citenamefont {Gepr{\"a}gs}\ \emph {et~al.}(2016)\citenamefont
  {Gepr{\"a}gs}, \citenamefont {Kehlberger}, \citenamefont {Coletta},
  \citenamefont {Qiu}, \citenamefont {Guo}, \citenamefont {Schulz},
  \citenamefont {Mix}, \citenamefont {Meyer}, \citenamefont {Kamra},
  \citenamefont {Althammer} \emph {et~al.}}]{geprags2016origin}%
  \BibitemOpen
  \bibfield  {author} {\bibinfo {author} {\bibfnamefont {S.}~\bibnamefont
  {Gepr{\"a}gs}}, \bibinfo {author} {\bibfnamefont {A.}~\bibnamefont
  {Kehlberger}}, \bibinfo {author} {\bibfnamefont {F.~D.}\ \bibnamefont
  {Coletta}}, \bibinfo {author} {\bibfnamefont {Z.}~\bibnamefont {Qiu}},
  \bibinfo {author} {\bibfnamefont {E.-J.}\ \bibnamefont {Guo}}, \bibinfo
  {author} {\bibfnamefont {T.}~\bibnamefont {Schulz}}, \bibinfo {author}
  {\bibfnamefont {C.}~\bibnamefont {Mix}}, \bibinfo {author} {\bibfnamefont
  {S.}~\bibnamefont {Meyer}}, \bibinfo {author} {\bibfnamefont
  {A.}~\bibnamefont {Kamra}}, \bibinfo {author} {\bibfnamefont
  {M.}~\bibnamefont {Althammer}}, \emph {et~al.},\ }\href@noop {} {\bibfield
  {journal} {\bibinfo  {journal} {Nature Communications}\ }\textbf {\bibinfo
  {volume} {7}},\ \bibinfo {pages} {1} (\bibinfo {year} {2016})}\BibitemShut
  {NoStop}%
\bibitem [{\citenamefont {Galanakis}\ \emph {et~al.}(2007)\citenamefont
  {Galanakis}, \citenamefont {{\"O}zdo{\u{g}}an}, \citenamefont
  {{\c{S}}a{\c{s}}{\i}o{\u{g}}lu},\ and\ \citenamefont
  {Akta{\c{s}}}}]{galanakis2007ab}%
  \BibitemOpen
  \bibfield  {author} {\bibinfo {author} {\bibfnamefont {I.}~\bibnamefont
  {Galanakis}}, \bibinfo {author} {\bibfnamefont {K.}~\bibnamefont
  {{\"O}zdo{\u{g}}an}}, \bibinfo {author} {\bibfnamefont {E.}~\bibnamefont
  {{\c{S}}a{\c{s}}{\i}o{\u{g}}lu}},\ and\ \bibinfo {author} {\bibfnamefont
  {B.}~\bibnamefont {Akta{\c{s}}}},\ }\href@noop {} {\bibfield  {journal}
  {\bibinfo  {journal} {Physical Review B}\ }\textbf {\bibinfo {volume} {75}},\
  \bibinfo {pages} {172405} (\bibinfo {year} {2007})}\BibitemShut {NoStop}%
\bibitem [{\citenamefont {Shi}\ \emph {et~al.}(2018)\citenamefont {Shi},
  \citenamefont {Muechler}, \citenamefont {Manna}, \citenamefont {Zhang},
  \citenamefont {Koepernik}, \citenamefont {Car}, \citenamefont {Van
  Den~Brink}, \citenamefont {Felser},\ and\ \citenamefont
  {Sun}}]{shi2018prediction}%
  \BibitemOpen
  \bibfield  {author} {\bibinfo {author} {\bibfnamefont {W.}~\bibnamefont
  {Shi}}, \bibinfo {author} {\bibfnamefont {L.}~\bibnamefont {Muechler}},
  \bibinfo {author} {\bibfnamefont {K.}~\bibnamefont {Manna}}, \bibinfo
  {author} {\bibfnamefont {Y.}~\bibnamefont {Zhang}}, \bibinfo {author}
  {\bibfnamefont {K.}~\bibnamefont {Koepernik}}, \bibinfo {author}
  {\bibfnamefont {R.}~\bibnamefont {Car}}, \bibinfo {author} {\bibfnamefont
  {J.}~\bibnamefont {Van Den~Brink}}, \bibinfo {author} {\bibfnamefont
  {C.}~\bibnamefont {Felser}},\ and\ \bibinfo {author} {\bibfnamefont
  {Y.}~\bibnamefont {Sun}},\ }\href@noop {} {\bibfield  {journal} {\bibinfo
  {journal} {Physical Review B}\ }\textbf {\bibinfo {volume} {97}},\ \bibinfo
  {pages} {060406} (\bibinfo {year} {2018})}\BibitemShut {NoStop}%
\bibitem [{\citenamefont {Patel}\ \emph {et~al.}(2018)\citenamefont {Patel},
  \citenamefont {Shinde}, \citenamefont {Gupta}, \citenamefont {Dabhi},\ and\
  \citenamefont {Jha}}]{patel2018first}%
  \BibitemOpen
  \bibfield  {author} {\bibinfo {author} {\bibfnamefont {P.~D.}\ \bibnamefont
  {Patel}}, \bibinfo {author} {\bibfnamefont {S.}~\bibnamefont {Shinde}},
  \bibinfo {author} {\bibfnamefont {S.~D.}\ \bibnamefont {Gupta}}, \bibinfo
  {author} {\bibfnamefont {S.~D.}\ \bibnamefont {Dabhi}},\ and\ \bibinfo
  {author} {\bibfnamefont {P.~K.}\ \bibnamefont {Jha}},\ }\href@noop {}
  {\bibfield  {journal} {\bibinfo  {journal} {Computational Condensed Matter}\
  }\textbf {\bibinfo {volume} {15}},\ \bibinfo {pages} {61} (\bibinfo {year}
  {2018})}\BibitemShut {NoStop}%
\bibitem [{\citenamefont {Ma}\ \emph {et~al.}(2017)\citenamefont {Ma},
  \citenamefont {Hegde}, \citenamefont {Munira}, \citenamefont {Xie},
  \citenamefont {Keshavarz}, \citenamefont {Mildebrath}, \citenamefont
  {Wolverton}, \citenamefont {Ghosh},\ and\ \citenamefont
  {Butler}}]{ma2017computational}%
  \BibitemOpen
  \bibfield  {author} {\bibinfo {author} {\bibfnamefont {J.}~\bibnamefont
  {Ma}}, \bibinfo {author} {\bibfnamefont {V.~I.}\ \bibnamefont {Hegde}},
  \bibinfo {author} {\bibfnamefont {K.}~\bibnamefont {Munira}}, \bibinfo
  {author} {\bibfnamefont {Y.}~\bibnamefont {Xie}}, \bibinfo {author}
  {\bibfnamefont {S.}~\bibnamefont {Keshavarz}}, \bibinfo {author}
  {\bibfnamefont {D.~T.}\ \bibnamefont {Mildebrath}}, \bibinfo {author}
  {\bibfnamefont {C.}~\bibnamefont {Wolverton}}, \bibinfo {author}
  {\bibfnamefont {A.~W.}\ \bibnamefont {Ghosh}},\ and\ \bibinfo {author}
  {\bibfnamefont {W.}~\bibnamefont {Butler}},\ }\href@noop {} {\bibfield
  {journal} {\bibinfo  {journal} {Physical Review B}\ }\textbf {\bibinfo
  {volume} {95}},\ \bibinfo {pages} {024411} (\bibinfo {year}
  {2017})}\BibitemShut {NoStop}%
\bibitem [{\citenamefont {Hames}\ and\ \citenamefont
  {Crangle}(1971)}]{hames1971ferromagnetism}%
  \BibitemOpen
  \bibfield  {author} {\bibinfo {author} {\bibfnamefont {F.}~\bibnamefont
  {Hames}}\ and\ \bibinfo {author} {\bibfnamefont {J.}~\bibnamefont
  {Crangle}},\ }\href@noop {} {\bibfield  {journal} {\bibinfo  {journal}
  {Journal of Applied Physics}\ }\textbf {\bibinfo {volume} {42}},\ \bibinfo
  {pages} {1336} (\bibinfo {year} {1971})}\BibitemShut {NoStop}%
\bibitem [{\citenamefont
  {Rodr{\'\i}guez-Carvajal}(1993)}]{rodriguez1993recent}%
  \BibitemOpen
  \bibfield  {author} {\bibinfo {author} {\bibfnamefont {J.}~\bibnamefont
  {Rodr{\'\i}guez-Carvajal}},\ }\href@noop {} {\bibfield  {journal} {\bibinfo
  {journal} {Physica B: Condensed Matter}\ }\textbf {\bibinfo {volume} {192}},\
  \bibinfo {pages} {55} (\bibinfo {year} {1993})}\BibitemShut {NoStop}%
\bibitem [{\citenamefont {Chakraborty}\ \emph {et~al.}(2022)\citenamefont
  {Chakraborty}, \citenamefont {Gupta}, \citenamefont {Pakhira}, \citenamefont
  {Choudhary}, \citenamefont {Biswas}, \citenamefont {Mudryk}, \citenamefont
  {Pecharsky}, \citenamefont {Johnson},\ and\ \citenamefont
  {Mazumdar}}]{chakraborty2022ground}%
  \BibitemOpen
  \bibfield  {author} {\bibinfo {author} {\bibfnamefont {S.}~\bibnamefont
  {Chakraborty}}, \bibinfo {author} {\bibfnamefont {S.}~\bibnamefont {Gupta}},
  \bibinfo {author} {\bibfnamefont {S.}~\bibnamefont {Pakhira}}, \bibinfo
  {author} {\bibfnamefont {R.}~\bibnamefont {Choudhary}}, \bibinfo {author}
  {\bibfnamefont {A.}~\bibnamefont {Biswas}}, \bibinfo {author} {\bibfnamefont
  {Y.}~\bibnamefont {Mudryk}}, \bibinfo {author} {\bibfnamefont {V.~K.}\
  \bibnamefont {Pecharsky}}, \bibinfo {author} {\bibfnamefont {D.~D.}\
  \bibnamefont {Johnson}},\ and\ \bibinfo {author} {\bibfnamefont
  {C.}~\bibnamefont {Mazumdar}},\ }\href@noop {} {\bibfield  {journal}
  {\bibinfo  {journal} {Physical Review B}\ }\textbf {\bibinfo {volume}
  {106}},\ \bibinfo {pages} {224427} (\bibinfo {year} {2022})}\BibitemShut
  {NoStop}%
\bibitem [{\citenamefont {Mazumdar}\ \emph {et~al.}(2000)\citenamefont
  {Mazumdar}, \citenamefont {Nagarajan}, \citenamefont {Gupta}, \citenamefont
  {Padalia},\ and\ \citenamefont {Vijayaraghavan}}]{mazumdar2000smni}%
  \BibitemOpen
  \bibfield  {author} {\bibinfo {author} {\bibfnamefont {C.}~\bibnamefont
  {Mazumdar}}, \bibinfo {author} {\bibfnamefont {R.}~\bibnamefont {Nagarajan}},
  \bibinfo {author} {\bibfnamefont {L.}~\bibnamefont {Gupta}}, \bibinfo
  {author} {\bibfnamefont {B.}~\bibnamefont {Padalia}},\ and\ \bibinfo {author}
  {\bibfnamefont {R.}~\bibnamefont {Vijayaraghavan}},\ }\href@noop {}
  {\bibfield  {journal} {\bibinfo  {journal} {Applied Physics Letters}\
  }\textbf {\bibinfo {volume} {77}},\ \bibinfo {pages} {895} (\bibinfo {year}
  {2000})}\BibitemShut {NoStop}%
\bibitem [{\citenamefont {Kroder}\ \emph {et~al.}(2019)\citenamefont {Kroder},
  \citenamefont {Manna}, \citenamefont {Kriegner}, \citenamefont {Sukhanov},
  \citenamefont {Liu}, \citenamefont {Borrmann}, \citenamefont {Hoser},
  \citenamefont {Gooth}, \citenamefont {Schnelle}, \citenamefont {Inosov} \emph
  {et~al.}}]{kroder2019spin}%
  \BibitemOpen
  \bibfield  {author} {\bibinfo {author} {\bibfnamefont {J.}~\bibnamefont
  {Kroder}}, \bibinfo {author} {\bibfnamefont {K.}~\bibnamefont {Manna}},
  \bibinfo {author} {\bibfnamefont {D.}~\bibnamefont {Kriegner}}, \bibinfo
  {author} {\bibfnamefont {A.}~\bibnamefont {Sukhanov}}, \bibinfo {author}
  {\bibfnamefont {E.}~\bibnamefont {Liu}}, \bibinfo {author} {\bibfnamefont
  {H.}~\bibnamefont {Borrmann}}, \bibinfo {author} {\bibfnamefont
  {A.}~\bibnamefont {Hoser}}, \bibinfo {author} {\bibfnamefont
  {J.}~\bibnamefont {Gooth}}, \bibinfo {author} {\bibfnamefont
  {W.}~\bibnamefont {Schnelle}}, \bibinfo {author} {\bibfnamefont {D.~S.}\
  \bibnamefont {Inosov}}, \emph {et~al.},\ }\href@noop {} {\bibfield  {journal}
  {\bibinfo  {journal} {Physical Review B}\ }\textbf {\bibinfo {volume} {99}},\
  \bibinfo {pages} {174410} (\bibinfo {year} {2019})}\BibitemShut {NoStop}%
\bibitem [{\citenamefont {Pakhira}\ \emph {et~al.}(2016)\citenamefont
  {Pakhira}, \citenamefont {Mazumdar}, \citenamefont {Ranganathan},
  \citenamefont {Giri},\ and\ \citenamefont {Avdeev}}]{pakhira2016large}%
  \BibitemOpen
  \bibfield  {author} {\bibinfo {author} {\bibfnamefont {S.}~\bibnamefont
  {Pakhira}}, \bibinfo {author} {\bibfnamefont {C.}~\bibnamefont {Mazumdar}},
  \bibinfo {author} {\bibfnamefont {R.}~\bibnamefont {Ranganathan}}, \bibinfo
  {author} {\bibfnamefont {S.}~\bibnamefont {Giri}},\ and\ \bibinfo {author}
  {\bibfnamefont {M.}~\bibnamefont {Avdeev}},\ }\href@noop {} {\bibfield
  {journal} {\bibinfo  {journal} {Physical Review B}\ }\textbf {\bibinfo
  {volume} {94}},\ \bibinfo {pages} {104414} (\bibinfo {year}
  {2016})}\BibitemShut {NoStop}%
\bibitem [{\citenamefont {Mydosh}(1993)}]{mydosh1993spin}%
  \BibitemOpen
  \bibfield  {author} {\bibinfo {author} {\bibfnamefont {J.~A.}\ \bibnamefont
  {Mydosh}},\ }\href@noop {} {\emph {\bibinfo {title} {Spin glasses: an
  experimental introduction}}}\ (\bibinfo  {publisher} {CRC Press},\ \bibinfo
  {year} {1993})\BibitemShut {NoStop}%
\bibitem [{\citenamefont {Pakhira}\ \emph {et~al.}(2020)\citenamefont
  {Pakhira}, \citenamefont {Sangeetha}, \citenamefont {Smetana}, \citenamefont
  {Mudring},\ and\ \citenamefont {Johnston}}]{pakhira2020ferromagnetic}%
  \BibitemOpen
  \bibfield  {author} {\bibinfo {author} {\bibfnamefont {S.}~\bibnamefont
  {Pakhira}}, \bibinfo {author} {\bibfnamefont {N.}~\bibnamefont {Sangeetha}},
  \bibinfo {author} {\bibfnamefont {V.}~\bibnamefont {Smetana}}, \bibinfo
  {author} {\bibfnamefont {A.-V.}\ \bibnamefont {Mudring}},\ and\ \bibinfo
  {author} {\bibfnamefont {D.~C.}\ \bibnamefont {Johnston}},\ }\href@noop {}
  {\bibfield  {journal} {\bibinfo  {journal} {Physical Review B}\ }\textbf
  {\bibinfo {volume} {102}},\ \bibinfo {pages} {024410} (\bibinfo {year}
  {2020})}\BibitemShut {NoStop}%
\bibitem [{\citenamefont {Gupta}\ \emph {et~al.}(2023)\citenamefont {Gupta},
  \citenamefont {Chakraborty}, \citenamefont {Pakhira}, \citenamefont {Biswas},
  \citenamefont {Mudryk}, \citenamefont {Kumar}, \citenamefont {Mukherjee},
  \citenamefont {Okram}, \citenamefont {Das}, \citenamefont {Pecharsky} \emph
  {et~al.}}]{gupta2023experimental}%
  \BibitemOpen
  \bibfield  {author} {\bibinfo {author} {\bibfnamefont {S.}~\bibnamefont
  {Gupta}}, \bibinfo {author} {\bibfnamefont {S.}~\bibnamefont {Chakraborty}},
  \bibinfo {author} {\bibfnamefont {S.}~\bibnamefont {Pakhira}}, \bibinfo
  {author} {\bibfnamefont {A.}~\bibnamefont {Biswas}}, \bibinfo {author}
  {\bibfnamefont {Y.}~\bibnamefont {Mudryk}}, \bibinfo {author} {\bibfnamefont
  {A.}~\bibnamefont {Kumar}}, \bibinfo {author} {\bibfnamefont
  {B.}~\bibnamefont {Mukherjee}}, \bibinfo {author} {\bibfnamefont {G.~S.}\
  \bibnamefont {Okram}}, \bibinfo {author} {\bibfnamefont {A.}~\bibnamefont
  {Das}}, \bibinfo {author} {\bibfnamefont {V.~K.}\ \bibnamefont {Pecharsky}},
  \emph {et~al.},\ }\href@noop {} {\bibfield  {journal} {\bibinfo  {journal}
  {Physical Review B}\ }\textbf {\bibinfo {volume} {107}},\ \bibinfo {pages}
  {184408} (\bibinfo {year} {2023})}\BibitemShut {NoStop}%
\bibitem [{\citenamefont {Kundu}\ \emph {et~al.}(2023)\citenamefont {Kundu},
  \citenamefont {Pakhira}, \citenamefont {Choudhary}, \citenamefont {Gupta},
  \citenamefont {Chakraborty}, \citenamefont {Lakshminarasimhan}, \citenamefont
  {Ranganathan}, \citenamefont {Johnson},\ and\ \citenamefont
  {Mazumdar}}]{MilyPRB}%
  \BibitemOpen
  \bibfield  {author} {\bibinfo {author} {\bibfnamefont {M.}~\bibnamefont
  {Kundu}}, \bibinfo {author} {\bibfnamefont {S.}~\bibnamefont {Pakhira}},
  \bibinfo {author} {\bibfnamefont {R.}~\bibnamefont {Choudhary}}, \bibinfo
  {author} {\bibfnamefont {S.}~\bibnamefont {Gupta}}, \bibinfo {author}
  {\bibfnamefont {S.}~\bibnamefont {Chakraborty}}, \bibinfo {author}
  {\bibfnamefont {N.}~\bibnamefont {Lakshminarasimhan}}, \bibinfo {author}
  {\bibfnamefont {R.}~\bibnamefont {Ranganathan}}, \bibinfo {author}
  {\bibfnamefont {D.~D.}\ \bibnamefont {Johnson}},\ and\ \bibinfo {author}
  {\bibfnamefont {C.}~\bibnamefont {Mazumdar}},\ }\href@noop {} {\bibfield
  {journal} {\bibinfo  {journal} {Phys. Rev. B}\ }\textbf {\bibinfo {volume}
  {107}},\ \bibinfo {pages} {094421} (\bibinfo {year} {2023})}\BibitemShut
  {NoStop}%
\bibitem [{\citenamefont {Pakhira}\ \emph {et~al.}(2018)\citenamefont
  {Pakhira}, \citenamefont {Mazumdar}, \citenamefont {Ranganathan},\ and\
  \citenamefont {Giri}}]{pakhira2018chemical}%
  \BibitemOpen
  \bibfield  {author} {\bibinfo {author} {\bibfnamefont {S.}~\bibnamefont
  {Pakhira}}, \bibinfo {author} {\bibfnamefont {C.}~\bibnamefont {Mazumdar}},
  \bibinfo {author} {\bibfnamefont {R.}~\bibnamefont {Ranganathan}},\ and\
  \bibinfo {author} {\bibfnamefont {S.}~\bibnamefont {Giri}},\ }\href@noop {}
  {\bibfield  {journal} {\bibinfo  {journal} {Physical Chemistry Chemical
  Physics}\ }\textbf {\bibinfo {volume} {20}},\ \bibinfo {pages} {7082}
  (\bibinfo {year} {2018})}\BibitemShut {NoStop}%
\bibitem [{\citenamefont {Kumar}\ \emph {et~al.}(2018)\citenamefont {Kumar},
  \citenamefont {Kaushik}, \citenamefont {Siruguri},\ and\ \citenamefont
  {Pandey}}]{kumar2018evidence}%
  \BibitemOpen
  \bibfield  {author} {\bibinfo {author} {\bibfnamefont {A.}~\bibnamefont
  {Kumar}}, \bibinfo {author} {\bibfnamefont {S.}~\bibnamefont {Kaushik}},
  \bibinfo {author} {\bibfnamefont {V.}~\bibnamefont {Siruguri}},\ and\
  \bibinfo {author} {\bibfnamefont {D.}~\bibnamefont {Pandey}},\ }\href@noop {}
  {\bibfield  {journal} {\bibinfo  {journal} {Physical Review B}\ }\textbf
  {\bibinfo {volume} {97}},\ \bibinfo {pages} {104402} (\bibinfo {year}
  {2018})}\BibitemShut {NoStop}%
\bibitem [{\citenamefont {Murayama}\ \emph {et~al.}(1986)\citenamefont
  {Murayama}, \citenamefont {Yokosawa}, \citenamefont {Miyako},\ and\
  \citenamefont {Wassermann}}]{murayama1986two}%
  \BibitemOpen
  \bibfield  {author} {\bibinfo {author} {\bibfnamefont {S.}~\bibnamefont
  {Murayama}}, \bibinfo {author} {\bibfnamefont {K.}~\bibnamefont {Yokosawa}},
  \bibinfo {author} {\bibfnamefont {Y.}~\bibnamefont {Miyako}},\ and\ \bibinfo
  {author} {\bibfnamefont {E.}~\bibnamefont {Wassermann}},\ }\href@noop {}
  {\bibfield  {journal} {\bibinfo  {journal} {Physical Review Letters}\
  }\textbf {\bibinfo {volume} {57}},\ \bibinfo {pages} {1785} (\bibinfo {year}
  {1986})}\BibitemShut {NoStop}%
\bibitem [{\citenamefont {Wang}\ \emph {et~al.}(2006)\citenamefont {Wang},
  \citenamefont {Bai}, \citenamefont {Pan}, \citenamefont {Wang} \emph
  {et~al.}}]{wang2006multiple}%
  \BibitemOpen
  \bibfield  {author} {\bibinfo {author} {\bibfnamefont {Y.~T.}\ \bibnamefont
  {Wang}}, \bibinfo {author} {\bibfnamefont {H.~Y.}\ \bibnamefont {Bai}},
  \bibinfo {author} {\bibfnamefont {M.~X.}\ \bibnamefont {Pan}}, \bibinfo
  {author} {\bibfnamefont {W.~H.}\ \bibnamefont {Wang}}, \emph {et~al.},\
  }\href@noop {} {\bibfield  {journal} {\bibinfo  {journal} {Physical Review
  B}\ }\textbf {\bibinfo {volume} {74}},\ \bibinfo {pages} {064422} (\bibinfo
  {year} {2006})}\BibitemShut {NoStop}%
\bibitem [{\citenamefont {Pakhira}\ \emph {et~al.}(2019)\citenamefont
  {Pakhira}, \citenamefont {Mazumdar}, \citenamefont {Avdeev}, \citenamefont
  {Bhowmik},\ and\ \citenamefont {Ranganathan}}]{pakhira2019spatially}%
  \BibitemOpen
  \bibfield  {author} {\bibinfo {author} {\bibfnamefont {S.}~\bibnamefont
  {Pakhira}}, \bibinfo {author} {\bibfnamefont {C.}~\bibnamefont {Mazumdar}},
  \bibinfo {author} {\bibfnamefont {M.}~\bibnamefont {Avdeev}}, \bibinfo
  {author} {\bibfnamefont {R.}~\bibnamefont {Bhowmik}},\ and\ \bibinfo {author}
  {\bibfnamefont {R.}~\bibnamefont {Ranganathan}},\ }\href@noop {} {\bibfield
  {journal} {\bibinfo  {journal} {Journal of Alloys and Compounds}\ }\textbf
  {\bibinfo {volume} {785}},\ \bibinfo {pages} {72} (\bibinfo {year}
  {2019})}\BibitemShut {NoStop}%
\bibitem [{\citenamefont {Samanta}\ \emph {et~al.}(2018)\citenamefont
  {Samanta}, \citenamefont {Bhobe}, \citenamefont {Das}, \citenamefont
  {Kumar},\ and\ \citenamefont {Nigam}}]{samanta2018reentrant}%
  \BibitemOpen
  \bibfield  {author} {\bibinfo {author} {\bibfnamefont {T.}~\bibnamefont
  {Samanta}}, \bibinfo {author} {\bibfnamefont {P.}~\bibnamefont {Bhobe}},
  \bibinfo {author} {\bibfnamefont {A.}~\bibnamefont {Das}}, \bibinfo {author}
  {\bibfnamefont {A.}~\bibnamefont {Kumar}},\ and\ \bibinfo {author}
  {\bibfnamefont {A.}~\bibnamefont {Nigam}},\ }\href@noop {} {\bibfield
  {journal} {\bibinfo  {journal} {Physical Review B}\ }\textbf {\bibinfo
  {volume} {97}},\ \bibinfo {pages} {184421} (\bibinfo {year}
  {2018})}\BibitemShut {NoStop}%
\bibitem [{\citenamefont {Mondal}\ \emph {et~al.}(2019)\citenamefont {Mondal},
  \citenamefont {Dan}, \citenamefont {Mondal}, \citenamefont {Bhowmik},
  \citenamefont {Ranganathan},\ and\ \citenamefont
  {Mazumdar}}]{mondal2019physical}%
  \BibitemOpen
  \bibfield  {author} {\bibinfo {author} {\bibfnamefont {B.}~\bibnamefont
  {Mondal}}, \bibinfo {author} {\bibfnamefont {S.}~\bibnamefont {Dan}},
  \bibinfo {author} {\bibfnamefont {S.}~\bibnamefont {Mondal}}, \bibinfo
  {author} {\bibfnamefont {R.}~\bibnamefont {Bhowmik}}, \bibinfo {author}
  {\bibfnamefont {R.}~\bibnamefont {Ranganathan}},\ and\ \bibinfo {author}
  {\bibfnamefont {C.}~\bibnamefont {Mazumdar}},\ }\href@noop {} {\bibfield
  {journal} {\bibinfo  {journal} {Physical Chemistry Chemical Physics}\
  }\textbf {\bibinfo {volume} {21}},\ \bibinfo {pages} {16923} (\bibinfo {year}
  {2019})}\BibitemShut {NoStop}%
\bibitem [{\citenamefont {Das}\ \emph {et~al.}(2003)\citenamefont {Das},
  \citenamefont {Paranjpe},\ and\ \citenamefont {Murayama}}]{das2003neutron}%
  \BibitemOpen
  \bibfield  {author} {\bibinfo {author} {\bibfnamefont {A.}~\bibnamefont
  {Das}}, \bibinfo {author} {\bibfnamefont {S.}~\bibnamefont {Paranjpe}},\ and\
  \bibinfo {author} {\bibfnamefont {S.}~\bibnamefont {Murayama}},\ }\href@noop
  {} {\bibfield  {journal} {\bibinfo  {journal} {Physica B: Condensed Matter}\
  }\textbf {\bibinfo {volume} {335}},\ \bibinfo {pages} {130} (\bibinfo {year}
  {2003})}\BibitemShut {NoStop}%
\bibitem [{\citenamefont {SK}\ \emph {et~al.}(1999)\citenamefont {SK},
  \citenamefont {Honda}, \citenamefont {Murayama},\ and\ \citenamefont
  {Tsuchiya}}]{sk1999neutron}%
  \BibitemOpen
  \bibfield  {author} {\bibinfo {author} {\bibfnamefont {D.~A.~P.}\
  \bibnamefont {SK}}, \bibinfo {author} {\bibfnamefont {S.}~\bibnamefont
  {Honda}}, \bibinfo {author} {\bibfnamefont {S.}~\bibnamefont {Murayama}},\
  and\ \bibinfo {author} {\bibfnamefont {Y.}~\bibnamefont {Tsuchiya}},\
  }\href@noop {} {\bibfield  {journal} {\bibinfo  {journal} {J. Phys. Condens.
  Matter}\ }\textbf {\bibinfo {volume} {11}},\ \bibinfo {pages} {5209}
  (\bibinfo {year} {1999})}\BibitemShut {NoStop}%
\bibitem [{\citenamefont {Halpern}\ and\ \citenamefont
  {Holstein}(1941)}]{halpern1941passage}%
  \BibitemOpen
  \bibfield  {author} {\bibinfo {author} {\bibfnamefont {O.}~\bibnamefont
  {Halpern}}\ and\ \bibinfo {author} {\bibfnamefont {T.}~\bibnamefont
  {Holstein}},\ }\href@noop {} {\bibfield  {journal} {\bibinfo  {journal}
  {Physical Review}\ }\textbf {\bibinfo {volume} {59}},\ \bibinfo {pages} {960}
  (\bibinfo {year} {1941})}\BibitemShut {NoStop}%
\bibitem [{\citenamefont {Mitsuda}\ \emph {et~al.}(1992)\citenamefont
  {Mitsuda}, \citenamefont {Yoshizawa},\ and\ \citenamefont
  {Endoh}}]{mitsuda1992neutron}%
  \BibitemOpen
  \bibfield  {author} {\bibinfo {author} {\bibfnamefont {S.}~\bibnamefont
  {Mitsuda}}, \bibinfo {author} {\bibfnamefont {H.}~\bibnamefont {Yoshizawa}},\
  and\ \bibinfo {author} {\bibfnamefont {Y.}~\bibnamefont {Endoh}},\
  }\href@noop {} {\bibfield  {journal} {\bibinfo  {journal} {Physical Review
  B}\ }\textbf {\bibinfo {volume} {45}},\ \bibinfo {pages} {9788} (\bibinfo
  {year} {1992})}\BibitemShut {NoStop}%
\bibitem [{\citenamefont {Kawamiya}\ \emph {et~al.}(1991)\citenamefont
  {Kawamiya}, \citenamefont {Nishino}, \citenamefont {Matsuo},\ and\
  \citenamefont {Asano}}]{kawamiya1991electrical}%
  \BibitemOpen
  \bibfield  {author} {\bibinfo {author} {\bibfnamefont {N.}~\bibnamefont
  {Kawamiya}}, \bibinfo {author} {\bibfnamefont {Y.}~\bibnamefont {Nishino}},
  \bibinfo {author} {\bibfnamefont {M.}~\bibnamefont {Matsuo}},\ and\ \bibinfo
  {author} {\bibfnamefont {S.}~\bibnamefont {Asano}},\ }\href@noop {}
  {\bibfield  {journal} {\bibinfo  {journal} {Physical Review B}\ }\textbf
  {\bibinfo {volume} {44}},\ \bibinfo {pages} {12406} (\bibinfo {year}
  {1991})}\BibitemShut {NoStop}%
\bibitem [{\citenamefont {Nishino}\ \emph {et~al.}(1993)\citenamefont
  {Nishino}, \citenamefont {Inoue}, \citenamefont {Asano},\ and\ \citenamefont
  {Kawamiya}}]{nishino1993anomalous}%
  \BibitemOpen
  \bibfield  {author} {\bibinfo {author} {\bibfnamefont {Y.}~\bibnamefont
  {Nishino}}, \bibinfo {author} {\bibfnamefont {S.-y.}\ \bibnamefont {Inoue}},
  \bibinfo {author} {\bibfnamefont {S.}~\bibnamefont {Asano}},\ and\ \bibinfo
  {author} {\bibfnamefont {N.}~\bibnamefont {Kawamiya}},\ }\href@noop {}
  {\bibfield  {journal} {\bibinfo  {journal} {Physical Review B}\ }\textbf
  {\bibinfo {volume} {48}},\ \bibinfo {pages} {13607} (\bibinfo {year}
  {1993})}\BibitemShut {NoStop}%
\bibitem [{\citenamefont {Kataoka}(2001)}]{kataoka2001resistivity}%
  \BibitemOpen
  \bibfield  {author} {\bibinfo {author} {\bibfnamefont {M.}~\bibnamefont
  {Kataoka}},\ }\href@noop {} {\bibfield  {journal} {\bibinfo  {journal}
  {Physical Review B}\ }\textbf {\bibinfo {volume} {63}},\ \bibinfo {pages}
  {134435} (\bibinfo {year} {2001})}\BibitemShut {NoStop}%
\bibitem [{\citenamefont {Samanta}\ \emph {et~al.}(2021)\citenamefont
  {Samanta}, \citenamefont {Srihari},\ and\ \citenamefont
  {Bhobe}}]{samanta2021itinerant}%
  \BibitemOpen
  \bibfield  {author} {\bibinfo {author} {\bibfnamefont {T.}~\bibnamefont
  {Samanta}}, \bibinfo {author} {\bibfnamefont {V.}~\bibnamefont {Srihari}},\
  and\ \bibinfo {author} {\bibfnamefont {P.~A.}\ \bibnamefont {Bhobe}},\
  }\href@noop {} {\bibfield  {journal} {\bibinfo  {journal} {Physica Status
  Solidi (b)}\ }\textbf {\bibinfo {volume} {258}},\ \bibinfo {pages} {2000461}
  (\bibinfo {year} {2021})}\BibitemShut {NoStop}%
\bibitem [{\citenamefont {Rossiter}(1991)}]{rossiter1991electrical}%
  \BibitemOpen
  \bibfield  {author} {\bibinfo {author} {\bibfnamefont {P.~L.}\ \bibnamefont
  {Rossiter}},\ }\href@noop {} {\emph {\bibinfo {title} {The electrical
  resistivity of metals and alloys}}},\ Vol.~\bibinfo {volume} {6}\ (\bibinfo
  {publisher} {Cambridge university press},\ \bibinfo {year}
  {1991})\BibitemShut {NoStop}%
\bibitem [{\citenamefont {Bombor}\ \emph {et~al.}(2013)\citenamefont {Bombor},
  \citenamefont {Blum}, \citenamefont {Volkonskiy}, \citenamefont {Rodan},
  \citenamefont {Wurmehl}, \citenamefont {Hess},\ and\ \citenamefont
  {B{\"u}chner}}]{bombor2013half}%
  \BibitemOpen
  \bibfield  {author} {\bibinfo {author} {\bibfnamefont {D.}~\bibnamefont
  {Bombor}}, \bibinfo {author} {\bibfnamefont {C.~G.}\ \bibnamefont {Blum}},
  \bibinfo {author} {\bibfnamefont {O.}~\bibnamefont {Volkonskiy}}, \bibinfo
  {author} {\bibfnamefont {S.}~\bibnamefont {Rodan}}, \bibinfo {author}
  {\bibfnamefont {S.}~\bibnamefont {Wurmehl}}, \bibinfo {author} {\bibfnamefont
  {C.}~\bibnamefont {Hess}},\ and\ \bibinfo {author} {\bibfnamefont
  {B.}~\bibnamefont {B{\"u}chner}},\ }\href@noop {} {\bibfield  {journal}
  {\bibinfo  {journal} {Physical Review Letters}\ }\textbf {\bibinfo {volume}
  {110}},\ \bibinfo {pages} {066601} (\bibinfo {year} {2013})}\BibitemShut
  {NoStop}%
\bibitem [{\citenamefont {Rani}\ \emph {et~al.}(2018)\citenamefont {Rani},
  \citenamefont {Kangsabanik}, \citenamefont {Suresh}, \citenamefont {Patra},
  \citenamefont {Bhattacharyya}, \citenamefont {Jha},\ and\ \citenamefont
  {Alam}}]{rani2018origin}%
  \BibitemOpen
  \bibfield  {author} {\bibinfo {author} {\bibfnamefont {D.}~\bibnamefont
  {Rani}}, \bibinfo {author} {\bibfnamefont {J.}~\bibnamefont {Kangsabanik}},
  \bibinfo {author} {\bibfnamefont {K.}~\bibnamefont {Suresh}}, \bibinfo
  {author} {\bibfnamefont {N.}~\bibnamefont {Patra}}, \bibinfo {author}
  {\bibfnamefont {D.}~\bibnamefont {Bhattacharyya}}, \bibinfo {author}
  {\bibfnamefont {S.}~\bibnamefont {Jha}},\ and\ \bibinfo {author}
  {\bibfnamefont {A.}~\bibnamefont {Alam}},\ }\href@noop {} {\bibfield
  {journal} {\bibinfo  {journal} {Physical Review Applied}\ }\textbf {\bibinfo
  {volume} {10}},\ \bibinfo {pages} {054022} (\bibinfo {year}
  {2018})}\BibitemShut {NoStop}%
\bibitem [{\citenamefont {{\'S}lebarski}\ \emph {et~al.}(2000)\citenamefont
  {{\'S}lebarski}, \citenamefont {Maple}, \citenamefont {Freeman},
  \citenamefont {Sirvent}, \citenamefont {Tworuszka}, \citenamefont
  {Orzechowska}, \citenamefont {Wrona}, \citenamefont {Jezierski},
  \citenamefont {Chiuzbaian},\ and\ \citenamefont
  {Neumann}}]{slebarski2000weak}%
  \BibitemOpen
  \bibfield  {author} {\bibinfo {author} {\bibfnamefont {A.}~\bibnamefont
  {{\'S}lebarski}}, \bibinfo {author} {\bibfnamefont {M.}~\bibnamefont
  {Maple}}, \bibinfo {author} {\bibfnamefont {E.}~\bibnamefont {Freeman}},
  \bibinfo {author} {\bibfnamefont {C.}~\bibnamefont {Sirvent}}, \bibinfo
  {author} {\bibfnamefont {D.}~\bibnamefont {Tworuszka}}, \bibinfo {author}
  {\bibfnamefont {M.}~\bibnamefont {Orzechowska}}, \bibinfo {author}
  {\bibfnamefont {A.}~\bibnamefont {Wrona}}, \bibinfo {author} {\bibfnamefont
  {A.}~\bibnamefont {Jezierski}}, \bibinfo {author} {\bibfnamefont
  {S.}~\bibnamefont {Chiuzbaian}},\ and\ \bibinfo {author} {\bibfnamefont
  {M.}~\bibnamefont {Neumann}},\ }\href@noop {} {\bibfield  {journal} {\bibinfo
   {journal} {Physical Review B}\ }\textbf {\bibinfo {volume} {62}},\ \bibinfo
  {pages} {3296} (\bibinfo {year} {2000})}\BibitemShut {NoStop}%
\bibitem [{\citenamefont {Bainsla}\ \emph {et~al.}(2014)\citenamefont
  {Bainsla}, \citenamefont {Suresh}, \citenamefont {Nigam}, \citenamefont
  {Manivel~Raja}, \citenamefont {Varaprasad}, \citenamefont {Takahashi},\ and\
  \citenamefont {Hono}}]{bainsla2014high}%
  \BibitemOpen
  \bibfield  {author} {\bibinfo {author} {\bibfnamefont {L.}~\bibnamefont
  {Bainsla}}, \bibinfo {author} {\bibfnamefont {K.}~\bibnamefont {Suresh}},
  \bibinfo {author} {\bibfnamefont {A.}~\bibnamefont {Nigam}}, \bibinfo
  {author} {\bibfnamefont {M.}~\bibnamefont {Manivel~Raja}}, \bibinfo {author}
  {\bibfnamefont {B.~C.~S.}\ \bibnamefont {Varaprasad}}, \bibinfo {author}
  {\bibfnamefont {Y.}~\bibnamefont {Takahashi}},\ and\ \bibinfo {author}
  {\bibfnamefont {K.}~\bibnamefont {Hono}},\ }\href@noop {} {\bibfield
  {journal} {\bibinfo  {journal} {Journal of Applied Physics}\ }\textbf
  {\bibinfo {volume} {116}},\ \bibinfo {pages} {203902} (\bibinfo {year}
  {2014})}\BibitemShut {NoStop}%
\bibitem [{\citenamefont {Lee}\ and\ \citenamefont
  {Ramakrishnan}(1985)}]{lee1985disordered}%
  \BibitemOpen
  \bibfield  {author} {\bibinfo {author} {\bibfnamefont {P.~A.}\ \bibnamefont
  {Lee}}\ and\ \bibinfo {author} {\bibfnamefont {T.}~\bibnamefont
  {Ramakrishnan}},\ }\href@noop {} {\bibfield  {journal} {\bibinfo  {journal}
  {Reviews of Modern Physics}\ }\textbf {\bibinfo {volume} {57}},\ \bibinfo
  {pages} {287} (\bibinfo {year} {1985})}\BibitemShut {NoStop}%
\end{thebibliography}
%

\end{document}